\documentclass[letterpaper,11pt]{article}

\usepackage{jheppub}
\usepackage{graphicx}
\usepackage{amsmath}
\usepackage{xspace}
\usepackage{xcolor}
\usepackage{cancel} 
\usepackage[normalem]{ulem}
\usepackage{bm}

\newcommand{\eq}[1]{eq.~\eqref{eq:#1}}
\newcommand{\eqs}[2]{eqs.~\eqref{eq:#1} and \eqref{eq:#2}}

\renewcommand{\sec}[1]{sec.~\ref{sec:#1}}

\newcommand{\subsec}[1]{sec.~\ref{subsec:#1}}

\newcommand{\fig}[1]{fig.~\ref{fig:#1}}

\newcommand{\app}[1]{app.~\ref{app:#1}}

\newcommand{\ord}[1]{\mathcal{O}(#1)}
\newcommand{\ca}[1]{\mathcal{#1}}

\newcommand{\be}{\begin{equation}}
\newcommand{\ee}{\end{equation}}

\newcommand{\df}{\mathrm{d}}
\newcommand{\img}{\mathrm{i}}

\newcommand{\e}{\epsilon}

\newcommand{\bn}{\bar{n}}
\newcommand{\bq}{{\bar{q}}}

\newcommand{\T}{\mathbf{T}}
\newcommand{\Sym}{\textrm{Sym}}
\newcommand{\bra}[1]{\langle #1 \rvert}
\newcommand{\ket}[1]{\lvert #1 \rangle}
\newcommand{\braket}[2]{\langle #1 \vert #2 \rangle}

\newcommand{\cA}{\mathcal{A}}

\newcommand{\cM}{\mathcal{M}}

\newcommand{\cT}{\mathcal{T}}
\newcommand{\cS}{\mathcal{S}}

\newcommand{\GeV}{\,\mathrm{GeV}}

\newcommand{\nn}{\nonumber}

\newcommand{\as}{\alpha_s}

\newcommand{\pTcut}{p_T^\mathrm{cut}}

\newcommand{\Li}{\textrm{Li}}

\newcommand{\wh}{\widehat}

\newcommand{\CS}{C\!S}
\newcommand{\Sp}{\textbf{Sp}}

\newcommand{\jet}{\mathrm{jet}}

\newcommand{\Split}{\mathrm{Split}}
\newcommand{\tree}{\mathrm{tree}}

\newcommand{\kt}{\mathrm{k}_\mathrm{T}}

\allowdisplaybreaks[4]

\title{Jet Veto Clustering Logarithms Beyond Leading Order}

\author[a]{Simone Alioli$^a$ and Jonathan R.~Walsh}

\affiliation[a]{Ernest Orlando Lawrence Berkeley National Laboratory, University of California, Berkeley, CA 94720, U.S.A.}

\emailAdd{salioli@lbl.gov}
\emailAdd{jwalsh@lbl.gov}

\abstract{
  
Many experimental analyses separate events into exclusive jet bins, using a jet algorithm to cluster the final state and then veto on jets.  Jet clustering induces logarithmic dependence on the jet radius $R$ in the cross section for exclusive jet bins, a dependence that is poorly controlled due to the non-global nature of the clustering.  At jet radii of experimental interest, the leading order (LO) clustering effects are numerically significant, but the higher order effects are currently unknown.  We rectify this situation by calculating the most important part of the next-to-leading order (NLO) clustering logarithms of $R$ for any 0-jet process, which enter as $\ord{\as^3}$ corrections to the cross section.  The calculation blends subtraction methods for NLO calculations with factorization properties of QCD and soft-collinear effective theory (SCET).  We compare the size of the known LO and new NLO clustering logarithms and find that the impact of the NLO terms on the 0-jet cross section in Higgs production is small.  This brings clustering effects under better control and may be used to improve uncertainty estimates on cross sections with a jet veto.

}

\begin{document}

%% fix date before submission
\maketitle

%%%%%%%%%%%%%%%%%%%%%%%%%%%%%%%%%%%%%%%%%%%%%%%%%%%%%%%%%%%%%%%%%%%%%%%%%%%%%%%%%%%%%%%%%%%%%%%%%%%%%%%%%
\section{Introduction}
\label{sec:intro}
%%%%%%%%%%%%%%%%%%%%%%%%%%%%%%%%%%%%%%%%%%%%%%%%%%%%%%%%%%%%%%%%%%%%%%%%%%%%%%%%%%%%%%%%%%%%%%%%%%%%%%%%%

An important ingredient in precision measurements of Higgs boson properties is the Standard Model prediction for cross sections used in the experimental analyses.  These analyses often involve cuts on the final state hadronic activity, vetoing candidate jets with transverse momentum $p_T$ above a veto scale $\pTcut$.  Exclusive jet cross sections, such as the $H + 0$-jet cross section in which there are no jets with $p_T > \pTcut$, play a key role in channels where the Higgs cannot be reconstructed and jet binning is an effective discriminant between various Standard Model backgrounds.  The uncertainty in theoretical predictions of these exclusive jet cross sections can be substantial, and accurately predicting these cross sections and providing robust estimates of their uncertainty is the focus of considerable effort \cite{Berger:2010xi,Stewart:2011cf,Banfi:2012yh,Becher:2012qa,Tackmann:2012bt,Banfi:2012jm,Liu:2012sz,Liu:2013hba,Becher:2013xia,Stewart:2013faa,Banfi:2013eda,Shao:2013uba}.  A good example is the $H \to WW^* \to \ell^+ \ell^- \nu \bar{\nu}$ channel, where a fixed order analysis of the exclusive 0-jet and 1-jet cross sections have theoretical uncertainties of $\ord{17\%}$ and $\ord{30\%}$ respectively and dominate the systematic uncertainties in the measurement \cite{ATLAS-CONF-2013-030,Chatrchyan:2013lba}.

There are two challenges in understanding cross sections with a jet veto.  The first is the fact that the jet veto scale, which is typically between 25 and 40 GeV, is well below the hard scales in the process, such as the Higgs mass $m_H$.  This leads to large logarithmic corrections of the ratio $\pTcut / m_H$.  Fixed order perturbation theory supplemented with resummation of these jet veto logarithms is a powerful tool to make predictions of exclusive jet cross sections and the associated uncertainties.  Recent work in this vein has substantially lowered the $H\to WW^*$ 0-jet and 1-jet uncertainties \cite{Banfi:2012jm,Liu:2013hba,Becher:2013xia,Stewart:2013faa,Banfi:2013eda}.

The second challenge is the effect of jet clustering.  Clustering from the jet algorithm leads to a complicated dependence on the jet radius $R$, a dependence that also takes part in and complicates the resummation of jet veto logarithms.  These effects start at $\ord{\as^2}$\footnote{The order counting of $\as$ is always relative to the LO cross section, which for Higgs production via gluon fusion is itself $\ord{\as^2}$.}, the first order in which there can be two final state partons and nontrivial clustering can take place.  At $\ord{\as^2}$ the clustering effects have been fully determined and have a substantial effect on the cross section \cite{Banfi:2012jm,Stewart:2013faa}.  No higher order terms are known, meaning the contribution of jet clustering to the cross section is effectively only known at lowest order.  It is challenging to make reliable uncertainty estimates of the higher order clustering effects from only the leading order term.  Calculating clustering contributions at the next order is the focus of this work.

We take the 0-jet cross section in Higgs production via gluon fusion as a test process, although results here can be reapplied to other color-singlet processes through a simple exchange of color factors.  The clustering effects have two types of terms: those proportional to logarithms of $m_H / \pTcut$ (``single logarithmic terms'') and terms independent of $\pTcut$ (``finite terms'').  At $\ord{\as^2}$ all of the clustering terms have been determined, and both single logarithmic and finite terms have contributions proportional to $\ln R$ \cite{Banfi:2012yh,Tackmann:2012bt}.

 In fact, the general form of the single logarithmic terms is known \cite{Tackmann:2012bt}: at each order, there is a new contribution multiplying the resummed cross section of the form
\begin{align} \label{eq:Undef}
U_{\rm clus}^{(n)} (R, \pTcut) = \exp \biggl[ \biggl(\frac{\as(\pTcut) C_A}{\pi}\biggr)^n C_n (R) \ln \frac{m_H}{\pTcut} \biggr] \,,
\end{align}
where
\begin{align}
\sigma_0 (\pTcut) = \sigma_0^{\rm sing} (\pTcut) + \sigma_0^{\rm ns} (\pTcut) \,, \qquad \sigma_0^{\rm sing} (\pTcut) \varpropto U_{\rm clus}^{(n)} (R, \pTcut) \,.
\end{align}
with $\sigma_0^{\rm sing}$ and $\sigma_0^{\rm ns}$ the singular and nonsingular 0-jet cross sections, respectively.  For phenomenological $\pTcut$ values, $\sigma_0^{\rm ns}$ is a small [$\ord{10\%}$] correction, meaning the impact on the cross section is almost directly proportional to $U_{\rm clus}^{(n)}$.

The exponentiation in \eq{Undef} comes from the resummation of the veto logarithm, and $C_n (R)$ is a function that can be decomposed in terms of logarithms of $R$ as
\be
C_n (R) = C_n^{(n-1)} \ln^{n-1} R^2 + \ldots + C_n^{(1)} \ln R^2 + C_n^{(0)} (R) \,,
\ee
where $C_n^{(0)} (R)$ is finite as $R\to0$.  Furthermore, the coefficients of the logarithms of $R$ can be connected to finite terms with logarithms of $R$ \cite{Stewart:2013faa}.  For a $q\bar{q}$ initiated color-singlet process, one factor of $C_A$ is replaced with a $C_F$ in \eq{Undef}.  

Standard jet radii used in experimental analyses are $R = 0.4$, $0.5$.  Formally, if one considers $R \sim \pTcut/m_H$ then the logarithms of $R$ are as important as the veto logarithms and resummation should be performed.  This implies that the leading coefficient $C_n^{(n-1)}$ at every order is part of a next-to-leading logarithmic (NLL) series.  It is not known if any of these coefficients are related, but if they are not then they have to be explicitly calculated at each order, rendering resummation\footnote{Here we want to distinguish between exponentiation and resummation.  Exponentiation involves capturing all higher-order terms (usually at a certain logarithmic order) that are generated by a given term, as in \eq{Undef}.  Resummation is more complex and involves capturing \textit{all} terms at a logarithmic order.  In the case here we can exponentiate a known $C_n^{(n-1)}$ to capture the NLL terms originating from it, but we cannot resum the NLL clustering logarithms unless all $C_n^{(n-1)}$ are known.} of the entire series impossible.  Note that the exponentiation of each term is known, as in \eq{Undef}.

At $\ord{\as^2}$, the leading clustering term has the coefficient \cite{Banfi:2012yh}
\begin{align} \label{eq:C21}
C_2^{(1)} &= \frac{1}{36} (131 - 12\pi^2 - 132\ln 2) + \frac{1}{18} (-23 + 24 \ln 2) \frac{T_F n_f}{C_A}\approx -2.49 \,.
\end{align}
For $\pTcut \in [25, 30]$ GeV and $R \in [0.4, 0.5]$, this term makes a contribution to $U_{\rm clus}^{(2)}$ which increases the cross section by an amount between 9\% and 15\%, which is on par with, or larger than, the theoretical uncertainties on the most recent predictions!  Clearly the higher order terms are important to not only provide precise predictions, but to understand the uncertainty associated to these clustering effects.

%%%%%%%%%%%%%%%%%%%%%%%%%%%%%%%%%%%%%%%%%%%%%%%%%%%%%%%%%%%%%%%%%%%%%%%%%%%%%%%%%%%%%%%%%%%%%%%%%%%%%%%%%
\subsection{Overview}
\label{subsec:overview}
%%%%%%%%%%%%%%%%%%%%%%%%%%%%%%%%%%%%%%%%%%%%%%%%%%%%%%%%%%%%%%%%%%%%%%%%%%%%%%%%%%%%%%%%%%%%%%%%%%%%%%%%%

In this work we calculate the coefficient $C_3^{(2)}$, which is the coefficient of the $\ln^2 R \, \ln m_H / \pTcut$ term in the $\ord{\as^3}$ correction to the cross section.  At first glance, the calculation of $C_3^{(2)}$ is very challenging as it naively requires a three loop calculation.  However, since the clustering effects start at $\ord{\as^2}$, the full coefficient $C_3 (R)$ is intrinsically an NLO quantity.  

There are three major simplifications that blend nicely in the calculation of $C_3^{(2)}$.  They are:
\begin{itemize}
\item \emph{Collinear factorization formulas that simplify the matrix elements relevant for the calculation of $C_3^{(2)}$}.  One can exploit the fact that the clustering logarithms arise from the region of phase space where final state particles are close together.  This means the leading clustering term at each order (given by the coefficient $C_n^{(n-1)}$) is determined by matrix elements with a collimated final state, allowing one to take advantage of powerful collinear factorization formulas \cite{Catani:1998nv,Catani:1999ss}.  This allows us to attack $C_3^{(2)}$ rather than the entire $C_3 (R)$.
\item \emph{Factorization properties of soft-collinear effective theory (SCET)} \cite{Bauer:2000ew, Bauer:2000yr, Bauer:2001ct, Bauer:2001yt, Bauer:2002nz}\emph{, which allow for a direct calculation of $C_3^{(2)}$ via the soft function}.  Factorization properties from SCET allow the fixed-order cross section to be factorized into beam and soft functions, where beam functions describe radiation along the beam directions (see Ref.~\cite{Stewart:2009yx}) and the soft function describes global radiation across the whole event.  This factorization allows one to cleanly separate the dynamics that gives rise to the $\ln m_H/\pTcut$ multiplying $C_3^{(2)}$ from the dynamics giving the logarithms of $R$.  The result is that $C_3^{(2)}$ can be framed as an NLO calculation in the soft function (one where the tree level contribution starts at $\ord{\as^2}$ in the soft function).
\item \emph{Subtractions for NLO calculations, which separately regulate the infrared divergent regions of phase space in the real and virtual contributions and allow one to perform the calculation numerically.}  Subtractions are an efficient method to handle the divergences present in real and virtual matrix elements.  These divergences cancel in the total result, but since the phase space of the real and virtual contributions are distinct, a numerical implementation of the calculation that implicitly sums these canceling divergences is challenging.  Subtractions separately render the real and virtual contributions finite by reproducing the divergences in the real matrix element in a way that can be analytically integrated to provide a cancellation of the divergences in the virtual matrix element.
\end{itemize}

The first two simplifications work in tandem to produce an NLO calculation that determines $C_3^{(2)}$.  However, the phase space restrictions required for the calculation make the singularity structure quite complex, and the matrix elements are still lengthy.  The calculation is made tractable by using subtractions.  Using the simplifications in the matrix element from QCD and SCET techniques, the number of singular regions of phase space is substantially reduced from a complete $H + 2$-jet NLO calculation.  When introducing the subtractions, we will highlight the connection between the matrix elements used for subtractions and the matrix elements of SCET.

In \sec{setup} we show how the calculation of $C_3 (R)$ can be reduced to a soft function calculation, and in \sec{splitting} we determine the matrix elements needed to calculate $C_3^{(2)}$ using splitting functions.  In \sec{subtractions} we set up the calculation of $C_3^{(2)}$ as an NLO calculation using FKS subtractions.  In \sec{virt} we determine the relevant terms in the virtual matrix elements needed for the calculation, as well as their combination with the integrated subtractions and the contribution to the result.  In \sec{calc} we perform the calculation of $C_3^{(2)}$, analyzing the impact on the $H + 0$-jet cross section and its uncertainties and discussing the all-orders series of clustering logarithms.  Finally, in \sec{conclusions} we give our outlook and conclude.

%%%%%%%%%%%%%%%%%%%%%%%%%%%%%%%%%%%%%%%%%%%%%%%%%%%%%%%%%%%%%%%%%%%%%%%%%%%%%%%%%%%%%%%%%%%%%%%%%%%%%%%%%
\section{Clustering Logarithms from the Soft Function}
\label{sec:setup}
%%%%%%%%%%%%%%%%%%%%%%%%%%%%%%%%%%%%%%%%%%%%%%%%%%%%%%%%%%%%%%%%%%%%%%%%%%%%%%%%%%%%%%%%%%%%%%%%%%%%%%%%%

The veto on jets found by a clustering algorithm can be viewed as a local veto, because the jet algorithm acts on local clusters of radiation.  This should be contrasted with a global jet veto, an event-wide measure that restricts jet activity that does not depend on a jet algorithm.  For example, the $H + 0$-jet cross section can be defined by a veto on beam thrust, and precision calculations can be carried out to high accuracy without the complications that arise in a clustering algorithm \cite{Berger:2010xi}.  Relevant to our case, the $E_T$ of the event is an effective global veto that is independent of the jet algorithm, where
\be
E_T = \sum_i p_{Ti} \,.
\ee
At $\ord{\as}$ there is only one particle in the final state and requiring $E_T < \pTcut$ is equivalent to the jet veto.  The difference
\be
\Delta \sigma (\pTcut) = \sigma (p_T^\jet < \pTcut) - \sigma (E_T < \pTcut)
\ee
is useful to understand clustering effects.  In the first term the jet veto is performed (where $p_T^\jet$ is the leading jet $p_T$), and in the second a global veto on $E_T$ is performed.  The global veto term is independent of the jet algorithm and any clustering effects, and may be used to understand the structure of the veto logarithms.  In this case it is helpful to isolate clustering effects, and hence the $R$-dependence, into the $\Delta\sigma$ clustering correction\footnote{The clustering correction defined by $\Delta\sigma$ depends on the choice of $E_T$ as the global veto (as opposed to, e.g., the Higgs $p_T$).  Since the global veto is independent of the algorithm, the clustering corrections should be viewed as the $R$-dependent terms plus a set of $R$-independent terms that depend on the choice of global veto.}.

Since the $\ord{\as}$ clustering correction vanishes, $\Delta \sigma^{(1)} (\pTcut) = 0$, the known $\ord{\as^2}$ clustering effects are a LO quantity.  This implies that $\Delta \sigma^{(3)} (\pTcut)$, which contains the \textit{complete} $C_3 (R)$, could be determined from existing $H + 2$-jet NLO codes.  However, $\Delta \sigma$ also contains higher powers of veto logarithms arising from exponentiation of lower order terms, and extracting $C_3 (R)$ requires an overwhelming computational investment (that is not very feasible) as it requires working in a small $R$ and $\pTcut$ regime to accurately extract the logarithmic dependence in $C_3 (R)$.  Therefore we opt to focus solely on the leading $\ln^2 R$ terms and calculate them through more direct methods.

In the small $R$ limit, the $H + 0$-jet cross section can be factorized in SCET into hard ($H$), beam ($B_{a,b}$), and soft ($S$) functions \cite{Becher:2012qa,Tackmann:2012bt}:
\be
\sigma (\pTcut) = H (m_H, \mu) \int \df Y \, B_a (\pTcut, m_H, x_a, \mu, \nu) \, B_b (\pTcut, m_H, x_b, \mu, \nu) \, S (\pTcut, \mu, \nu) \,.
\ee
The bare soft and beam functions individually contain rapidity divergences that are not regulated by dimensional regularization.  These rapidity divergences can be regulated in different ways, and in this work we use the rapidity renormalization group \cite{Chiu:2011qc,Chiu:2012ir}.  This method regulates the rapidity divergences with an explicit factor in matrix elements, introducing a scale $\nu$ that functions much like $\mu$ in dimensional regularization.

In SCET with the rapidity regulator the clustering logarithms from $C_n (R)$ are divided between the beam and soft functions.  At $\ord{\as^n}$, the contributions from $C_n (R)$ are \cite{Tackmann:2012bt}
\begin{align} \label{eq:Cncontributions}
\textrm{total} \; & \varpropto \; \Bigl( \frac{\as}{4\pi} \Bigr)^n C_n (R) \ln \frac{m_H}{\pTcut} \,, \nn \\
\textrm{soft} \; & \varpropto \; \Bigl( \frac{\as}{4\pi} \Bigr)^n C_n (R) \ln \frac{\nu}{\pTcut} \,, \nn \\
\textrm{beam} \; & \varpropto \; \Bigl( \frac{\as}{4\pi} \Bigr)^n C_n (R) \ln \frac{m_H}{\nu} \,.
\end{align}
Therefore $C_n (R)$ can be calculated from the soft function, simplifying the computation.

%%%%%%%%%%%%%%%%%%%%%%%%%%%%%%%%%%%%%%%%%%%%%%%%%%%%%%%%%%%%%%%%%%%%%%%%%%%%%%%%%%%%%%%%%%%%%%%%%%%%%%%%%
\subsection{The Soft Function}
\label{subsec:softcalc}
%%%%%%%%%%%%%%%%%%%%%%%%%%%%%%%%%%%%%%%%%%%%%%%%%%%%%%%%%%%%%%%%%%%%%%%%%%%%%%%%%%%%%%%%%%%%%%%%%%%%%%%%%

The soft function in SCET is a forward scattering matrix element of soft Wilson lines with a veto measurement on the final state,
\be
S (\pTcut, \mu, \nu) = \bigl\langle 0 \bigl\lvert Y^{\dag}_{\bn} \, Y_n \, \cM(\pTcut) \, Y^{\dag}_n \, Y_{\bn} \bigr\rvert 0 \bigr\rangle \,.
\ee
The $Y_n$ are soft Wilson lines, and $n = (1,0,0,1),\,\bn = (1,0,0,-1)$ are light-like vectors oriented along the beam direction.  The measurement function for the veto cross section is
\be
\cM(\pTcut) = \prod_{\textrm{jets } j} \theta(p_{Tj} < \pTcut) \,,
\ee
where the jets are formed by clustering the soft particles in the final state.  For the clustering correction (relative to the $E_T$ global veto), the measurement function is
\be
\Delta \cM(\pTcut) = \prod_{\textrm{jets } j} \theta(p_{Tj} < \pTcut) - \theta \Bigl( \sum_i p_{Ti} < \pTcut \Bigr) \,,
\ee
which defines a soft function correction $\Delta S$.  At $\ord{\as}$, $\Delta \cM$ vanishes, and at $\ord{\as^2}$ and $\ord{\as^3}$ in the small $R$ limit\footnote{In the small $R$ limit, we can replace the vector sum over the momentum in each jet with the scalar sum; the difference is power suppressed in $R$.} it is
\begin{align} \label{eq:DeltaM23}
&\Delta \cM^{(2)} (\pTcut) = \theta(\Delta R_{12} > R) \bigl[ \theta(p_{T1} < \pTcut) \theta(p_{T2} < \pTcut) - \theta(p_{T1} + p_{T2} < \pTcut) \bigr] \,, \\
&\Delta \cM^{(3)} (\pTcut) = \theta(\textrm{3-jet}) \bigl[ \theta(p_{T1} < \pTcut) \theta(p_{T2} < \pTcut) \theta(p_{T3} < \pTcut) - \theta(p_{T1} + p_{T2} + p_{T3} < \pTcut) \bigr] \nn \\
& \qquad + \theta(\textrm{2-jet};\,\{1+2,3\}) \bigl[ \theta(p_{T1} + p_{T2} < \pTcut) \theta(p_{T3} < \pTcut) - \theta(p_{T1} + p_{T2} + p_{T3} < \pTcut) \bigr] \nn \\
& \qquad + \theta(\textrm{2-jet};\,\{1+3,2\}) \bigl[ \theta(p_{T1} + p_{T3} < \pTcut) \theta(p_{T2} < \pTcut) - \theta(p_{T1} + p_{T2} + p_{T3} < \pTcut) \bigr] \nn \\
& \qquad + \theta(\textrm{2-jet};\,\{2+3,1\}) \bigl[ \theta(p_{T2} + p_{T3} < \pTcut) \theta(p_{T1} < \pTcut) - \theta(p_{T1} + p_{T2} + p_{T3} < \pTcut) \bigr] \,. \nn
\end{align}
The $p_{Ti}$ are the transverse momenta of the different particles, and at $\ord{\as^3}$ the outcomes of the jet algorithm are classified by constraints of the form $\theta(\textrm{2-jet} ;\, \{1+2,3\})$, which for example requires the algorithm to yield two jets, one with particles 1 and 2 and the other with particle 3.  These are jet algorithm dependent, and here we use the $\kt$-type clustering algorithms ($\kt$, Cambridge/Aachen, and anti-$\kt$ \cite{Catani:1991hj,Catani:1993hr,Ellis:1993tq,Dokshitzer:1997in,Cacciari:2008gp}), which allow us to study the algorithm dependence of the coefficient we calculate.  The $\ord{\as^2}$ phase space constraint depends only on the separation $\Delta R_{12} = \sqrt{\Delta y_{12}^2 + \Delta \phi_{12}^2}$ between the final state particles, and is common between $\kt$-type clustering algorithms.  The constraint $\theta(\Delta R_{12} > R)$ requires the two particles to be in different jets.  A study of the singular limits in these measurement functions shows that $\Delta \cM^{(2)}$ vanishes in the soft or collinear limits, and $\Delta \cM^{(3)}$ is nonvanishing only if a single parton becomes soft, a pair of partons become collinear, or in the combined soft-collinear limit.  Hence the calculation is LO at $\ord{\as^2}$ and NLO at $\ord{\as^3}$.  From the expressions in \eq{DeltaM23}, one sees that the measurement function vanishes if either of the following is true:
\be
\Delta\cM^{(n)} = 0 \textrm{ if } \sum_i p_{Ti} < \pTcut \textrm{ or any } p_{Ti} > \pTcut \,.
\ee
These serve as useful bounds on the phase space.  We note that this discussion is independent of the soft function, and also applies when considering the full QCD calculation.

The soft function is invariant under boosts along and rotations around the beam direction, suggesting a certain set of phase space variables.  Each on-shell (final state) phase space integral can be expressed in terms of the transverse momentum $p_T$, the rapidity $y$, and the azimuthal angle $\phi$.  In $d = 4-2\e$ dimensions,
\begin{align}
\int \df\Phi &\equiv \int \frac{\df^d p}{(2\pi)^d} 2\pi \delta(p^2) \theta(p^0) \nn \\
&= \frac{2}{(4\pi)^2} \frac{(4\pi)^\e}{\Gamma(1-\e)} \int_0^{\infty} \df p_T \, p_T^{1-2\e} \int_{-\infty}^{\infty} \df y \int_{-\pi}^{\pi} \frac{\df \phi}{2\pi} \sin^{-2\e} \phi \, \frac{\pi^{1/2}\Gamma(1 - \e)}{\Gamma(1/2 - \e)} \,.
\end{align}
Since the matrix elements and measurement are independent of the total rapidity ($y_t$) and azimuthal angle ($\phi_t$), a useful combination of the above variables is the following,
\begin{align}
x_i &\equiv \frac{p_{Ti}}{\pTcut} \,, & i &= 1,\, \ldots, n \,, \nn \\
y_{1i} &\equiv y_1 - y_i \,, \quad \phi_{1i} \equiv \phi_1 - \phi_i \,, & i &= 2,\, \ldots, n \,, \nn \\
y_t &\equiv \frac{1}{n} \sum_i y_i \,, \quad \phi_t \equiv \frac{1}{n} \sum_i \phi_i \,.
\end{align}
Note that it is also possible to define a dimensionful variable for the total $p_T$ and rescale each transverse momentum by it; since this is the only dimensionful phase space variable its dependence in the matrix element is fixed.  We will find use for these variables, and define
\be
p_{Tt} \equiv p_{T1} + \ldots + p_{Tn} \,, \qquad z_i \equiv \frac{p_{Ti}}{p_{Tt}} \,.
\ee
For the phase space integration we find it is more useful to rescale by the fixed $\pTcut$ and keep the $x_i$.  The phase space in terms of our chosen variables is
\begin{align}
\int \df\Phi_n &= \prod_{i=1}^n \frac{2}{(4\pi)^2} \frac{(4\pi)^\e}{\Gamma(1-\e)} \int_0^{\infty} p_{Ti}^{1-2\e} \df p_{Ti} \int_{-\infty}^{\infty} \df y_i \int_{-\pi}^{\pi} \frac{\df \phi_i}{2\pi} \, c_{\phi i} \nn \\
&= \biggl\{ \frac{2}{(4\pi)^2} \frac{(4\pi)^\e}{\Gamma(1-\e)} \, (\pTcut)^{2n-2n\e} \int_{-\infty}^{\infty} \df y_t \int_{-\pi}^{\pi} \frac{\df\phi_t}{2\pi} \, c_{\phi t} \biggr\} \nn \\
& \qquad \times \biggl\{ \biggl( \prod_{i=1}^n \int_0^{\infty} x_i^{1-2\e} \df x_i \biggr) \biggl( \prod_{i = 2}^n \frac{2}{(4\pi)^2} \frac{(4\pi)^\e}{\Gamma(1-\e)} \int_{-\infty}^{\infty} \df y_{1i} \int_{-\pi}^{\pi} \frac{\df\phi_{1i}}{2\pi} \, c_{\phi 1i} \biggr) \biggr\} \nn \\
&\equiv \biggl\{ \int \df\Phi_t \biggr\} \biggl\{ \int \df\wh\Phi_n \biggr\} \,,
\end{align}
with
\begin{align}
c_\phi \equiv [\pi^{1/2} \Gamma(1-\e) / \Gamma(1/2 - \e)] \sin^{-2\e} \phi \,.
\end{align}
For those not familiar with soft functions it is worthwhile to note that an arbitrary amount of momentum can go into the final state, as evidenced by the fact that there is no momentum conservation relation present in the phase space.

%%%%%%%%%%%%%%%%%%%%%%%%%%%%%%%%%%%%%%%%%%%%%%%%%%%%%%%%%%%%%%%%%%%%%%%%%%%%%%%%%%%%%%%%%%%%%%%%%%%%%%%%%
\subsection{Projecting Out the Veto Logarithm}
\label{subsec:parentPS}
%%%%%%%%%%%%%%%%%%%%%%%%%%%%%%%%%%%%%%%%%%%%%%%%%%%%%%%%%%%%%%%%%%%%%%%%%%%%%%%%%%%%%%%%%%%%%%%%%%%%%%%%%

Consider the soft function contribution at a given order, with the $d$-dimensional, $\ell$-loop matrix element denoted $\cT_{a_1 \ldots a_n}^{(\ell)} (\Phi_n)$, where all spin and color have been summed over and the $a_i$ serve as final state parton labels.  We only include the single c-web terms in the matrix element, as these are precisely the terms contributing to $C_n (R)$ \cite{Frenkel:1984pz,Tackmann:2012bt}.  Note that the number of parton labels $a_i$ denote how many particles are in the final state (the remaining are loop momenta which are integrated over).  The bare soft function clustering correction contribution from these matrix elements is
\be
\Delta S^{(n)} (\pTcut) = \int \df\Phi_n \sum_{k=0}^n \sum_{a_1, \ldots, a_k} \Sym_{a_1 \cdots a_k} \cT_{a_1 \ldots a_k}^{(n-k)} (\Phi_k) \Delta \cM^{(n)} (\pTcut, \Phi_k) \,,
\ee
where $\Sym_{a_1 \ldots a_k}$ is the symmetry factor for the particular channel.  The only non-unit symmetry factors relevant to this calculation are $\Sym_{gg} = 1/2!$ and $\Sym_{ggg} = 1/3!$.  For simplicity, we first consider more detailed properties of the fully real matrix element, $\cT_n^{(0)}$.  We can make the dependence of the matrix element on the dimensional regularization scale $\mu$, the coupling, and the parent phase space explicit:
\be \label{eq:Thatdef}
\cT_{a_1 \ldots a_n}^{(0)} (\Phi_n) = (4\pi\as \mu^{2\e})^n (\pTcut)^{-2n} \, \wh{\cT}_{a_1 \ldots a_n}^{(0)} (\wh\Phi_n) \,.
\ee
The measurement function only depends on $\wh\Phi_n$.

One notices that neither the matrix element nor the measurement function depends on $y_t$ and $\phi_t$ variables in the parent phase space, and it seems we can integrate over them with impunity.  However, since the $y_t$ integral spans an infinite range, we get an unregulated divergence.  This is a rapidity divergence, and there are canceling rapidity divergences in the soft and collinear sectors that must each be regulated.  The rapidity renormalization group provides a regularization scheme to regulate rapidity divergences and renormalize them much in the same way that dimensional regularization regulates virtuality divergences \cite{Chiu:2011qc,Chiu:2012ir}.  Resummation of rapidity logarithms using the rapidity renormalization group follows familiar steps to conventional resummation.  For a single c-web, the regulator factor in the soft function is
\be
R_{\eta} = \nu^{\eta} \lvert p_{3g} \rvert^{-\eta} \,,
\ee
where $p_{3g}$ is the component of the group momentum for the c-web along the Wilson line direction (the beam direction).  Since we only deal with a single c-web, the group momentum is the sum of all momenta in $\Phi_n$.  For our purposes, we only need the leading singularity:
\be
\int_{-\infty}^{\infty} \df y_t \, R_{\eta} = \nu^{\eta} p_{Tt}^{-\eta} \, \frac{2}{\eta} + \ord{\eta^0} \,.
\ee
This leading divergence produces a finite term proportional to $\ln \nu/\pTcut$, which is the logarithm that appears in \eq{Cncontributions}.  It is useful to note that the $\ord{\eta^0}$ term contains kinematic factors that regulate the combined collinear limit of the c-web (which implies it will only contribute to subleading clustering logarithms).

Integrating over $\Phi_t$, the result is
\begin{align}
\Delta S_{n}^{(n)} (\pTcut) &= \frac{2}{(4\pi)^2} \frac{(4\pi)^\e}{\Gamma(1-\e)} (4\pi\as)^n \biggl(\frac{\mu}{\pTcut}\biggr)^{2n\e} \biggl(\frac{\nu}{\pTcut}\biggr)^\eta \, \frac{2}{\eta} \nn \\
& \qquad \times \int \df\wh\Phi_n \biggl(\frac{p_{Tt}}{\pTcut}\biggr)^{-\eta} \sum_{a_1, \ldots, a_n} \Sym_{a_1 \ldots a_n} \wh\cT_{a_1 \ldots a_n}^{(0)} (\wh\Phi_n) \Delta \cM^{(n)} (\pTcut, \wh\Phi_n) \,.
\end{align}
At $\ord{\as^2}$ and $\ord{\as^3}$, this is
\begin{align} \label{eq:bornrealSn}
\Delta S_2^{(2)} (\pTcut) &= \Bigl( \frac{\as}{\pi} \Bigr)^2 \frac12 \frac{e^{2\gamma_E\e}}{\Gamma(1-\e)^2} \biggl( \frac{\mu}{\pTcut} \biggr)^{4\e} \frac{1}{\eta} \biggl(\frac{\nu}{\pTcut}\biggr)^\eta \int_0^\infty \df x_1 \df x_2 (x_1 x_2)^{1-2\e} (x_1 + x_2)^{-\eta} \nn \\
& \qquad \times \int_{-\infty}^{\infty} \df y \int_{-\pi}^{\pi} \frac{\df\phi}{2\pi} \, c_\phi \, \sum_{a_1, a_2} \Sym_{a_1 a_2} \wh{\cT}^{(0)}_{a_1 a_2} (\wh{\Phi}_2) \Delta \cM^{(2)} (\pTcut, \wh{\Phi}_2) \,, \nn \\
\Delta S_3^{(3)} (\pTcut) &= \Bigl( \frac{\as}{\pi} \Bigr)^3 \frac14 \frac{e^{3\gamma_E\e}}{\Gamma(1-\e)^3} \biggl( \frac{\mu}{\pTcut} \biggr)^{6\e} \frac{1}{\eta} \biggl(\frac{\nu}{\pTcut}\biggr)^\eta \int_0^\infty \df x_1 \df x_2 \df x_3 (x_1 x_2 x_3)^{1-2\e} \nn \\
& \qquad \times (x_1 + x_2 + x_3)^{-\eta} \int_{-\infty}^{\infty} \df y_{12} \, \df y_{13} \int_{-\pi}^{\pi} \frac{\df \phi_{12}}{2\pi} \frac{\df \phi_{13}}{2\pi} \, c_{\phi12} \, c_{\phi13} \nn \\
& \qquad \times \sum_{a_1, a_2, a_3} \Sym_{a_1 a_2 a_3} \wh{\cT}^{(0)}_{a_1 a_2 a_3} (\wh{\Phi}_3) \Delta \cM^{(3)} (\wh{\Phi}_3) \,.
\end{align}
Note that there is no additional divergence in the rapidity regulator parameter $\eta$ besides the overall $1/\eta$ in \eq{bornrealSn}.  If we are only concerned with the term containing the logarithm of $\pTcut$, we can expand in $\eta$ and obtain the coefficient of $\ln \nu/\pTcut$, which is directly related to $C_n (R)$ [see \eq{Cncontributions}].  There are remaining divergences in $\e$, though.  The measurement function cuts off the ultraviolet region of phase space, meaning they are infrared singularities which cancel between real and virtual contributions.

%%%%%%%%%%%%%%%%%%%%%%%%%%%%%%%%%%%%%%%%%%%%%%%%%%%%%%%%%%%%%%%%%%%%%%%%%%%%%%%%%%%%%%%%%%%%%%%%%%%%%%%%%
\section{The Real Matrix Elements in Terms of Splitting Functions}
\label{sec:splitting}
%%%%%%%%%%%%%%%%%%%%%%%%%%%%%%%%%%%%%%%%%%%%%%%%%%%%%%%%%%%%%%%%%%%%%%%%%%%%%%%%%%%%%%%%%%%%%%%%%%%%%%%%%

\begin{figure}[ht!]
\begin{center}
\includegraphics[width=0.45\columnwidth]{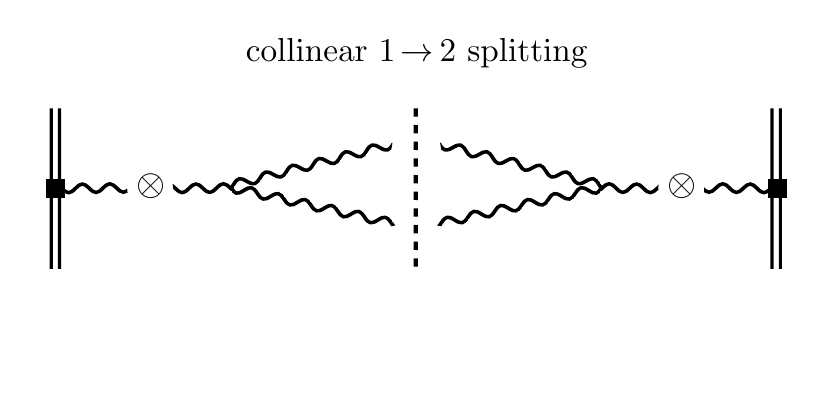} \hspace{3ex} \vspace{-4ex} \\
\includegraphics[width=0.45\columnwidth]{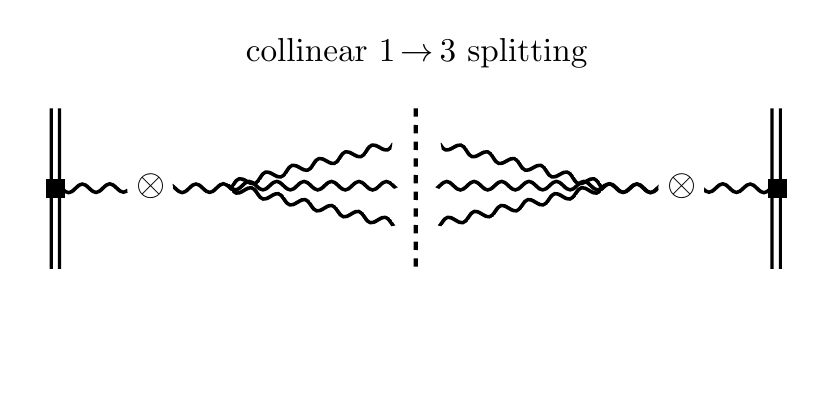} \hspace{3ex}
\includegraphics[width=0.45\columnwidth]{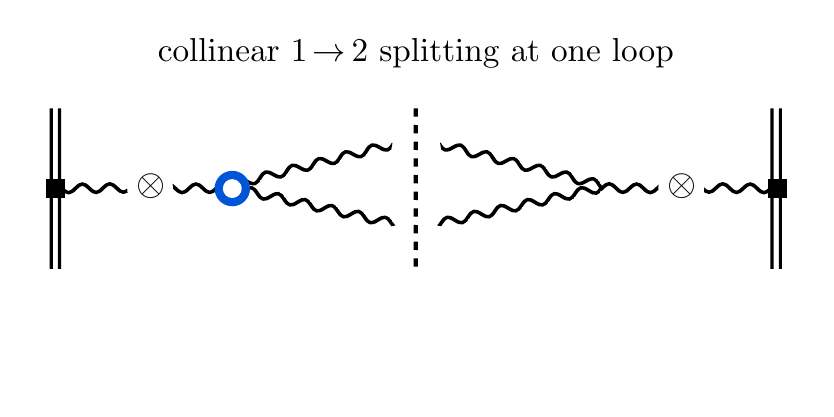} \hspace{3ex} \vspace{-4ex}
\caption{Schematic form of the leading order, real emission, and virtual matrix elements.  In each case the parent gluon is emitted from the soft Wilson line (the double lines) and undergoes a collinear splitting.  The collinear factorization is represented by `$\otimes$', with the splitting functions giving the matrix element in terms of the final state particles (those crossing the dashed line).  In the virtual matrix elements, the loop contribution is shown in blue.}
\label{fig:splitting}
\end{center}
\end{figure}

The leading clustering logarithm dependence [$\ln^2 R$ at $\ord{\as^3}$] arises from the collimated limit of the final state.  This implies that if we want to calculate $C_3^{(2)}$ alone, we can use matrix elements in this limit.  At tree level, these matrix elements can be built from lower order amplitudes by exploiting collinear factorization of tree-level amplitudes \cite{Ellis:1978sf,Ellis:1978ty,Campbell:1997hg,Catani:1998nv,Catani:1999ss}.  This factorization manifests at the lowest level in terms of the factorization of color-ordered amplitudes for individual helicity configurations.  At a less exclusive level, this factorization can be phrased in terms of splitting functions.  In \fig{splitting} we give a schematic picture of the matrix elements.
%%%
%%%

We note that to calculate the single $\ln R$ terms at $\ord{\as^3}$, whose coefficient is $C_3^{(1)}$, the matrix elements outside of the fully collimated limit are needed.  In general, the number of logarithms of $R$ is given by the number of collinear propagators in the matrix element \cite{Tackmann:2012bt}, meaning that the $\ln R$ terms arise from only one pair of partons becoming collinear (instead of all of them).  Since the $\ord{\as^3}$ matrix elements used here are in the triple collinear limit, they will not capture the complete $C_3^{(1)}$.

We adopt the basic notation in \cite{Catani:1998nv,Catani:1999ss}.  If $\cA_{a_1 \ldots}^{c_1, \ldots ; s_1, \ldots} (p_1, \ldots)$ is the tree-level amplitude to produce momenta $p_i$ with parton flavors $a_i$, colors $c_i$, and spins $s_i$, then the matrix element squared, summed over color and spin, is
\be
\cT^{(0)}_{a_1 \ldots} (p_1, \ldots) = \sum_{c_i, s_i} \big\lvert \cA_{a_1 \ldots}^{c_1, \ldots ; s_1, \ldots} (p_1, \ldots) \big\rvert^2 \,.
\ee
If a set of momenta $\{p_1, \ldots, p_n\}$ become collinear, then the matrix element factorizes,
\be \label{eq:collfact}
\cT^{(0)}_{a_1 \ldots a_n \ldots} (p_1, \ldots, p_n, \ldots) = \frac{2^{n-1} (4\pi\as\mu^{2\e})^{n-1}}{s_{1\ldots n}^{n-1}} \, \cT^{(0)\, ss'}_{a,\ldots} (p, \ldots) \hat{P}_{a_1 \ldots a_n}^{ss'} \,.
\ee
The momentum $p$ is the collinear limit of $p_1 + \cdots + p_n$ with flavor $a$.  The spin-polarization tensor $\cT^{ss'}$ is obtained by summing over all other spins,
\be
\cT^{(0)\, ss'}_{a\ldots} (p, \ldots) = \sum_{{\rm spins}\neq s,s'} \sum_{\rm colors} \bigl[ \cA^{c, \ldots; s,\ldots}_{a\ldots} (p,\ldots) \bigr] \bigl[ \cA^{c, \ldots; s',\ldots}_{a\ldots} (p,\ldots) \bigr]^{\dag} \,.
\ee
The momenta $p_i$ can be decomposed in the collinear limit as
\be
p_i^\mu = z_i p^\mu + k_{\perp i}^\mu - \frac{k_{\perp i}^2}{z_i} \frac{\bn_p^\mu}{2\bn_p \cdot p} \,,
\ee
where $\bn_p = (1, -\hat{p})$ is a null vector and the component of each momentum transverse to the collinear direction is
\begin{align}
k_{\perp i}^{\mu} &= z_i p_{Tt} \bigl[ \delta y_i \, y_{\perp t}^\mu + \delta \phi_i \, \phi_{\perp t}^\mu \bigr] \,, \nn \\
\delta y_i &\equiv y_i - \sum_j z_j y_j \,, & \delta \phi_i &\equiv \phi_i - \sum_j z_j \phi_j \,, \nn \\
y_{\perp t}^\mu &\equiv (\sinh y_t, 0, 0, \cosh y_t) \,, & \phi_{\perp t}^{\mu} &\equiv (0, -\sin\phi_t, \cos\phi_t, 0) \,.
\end{align}
Note that this decomposition satisfies $\sum_i z_i = 1$ and $\sum_i k_{\perp i}^\mu = 0$.

In our case, we are interested in the entire final state being collimated.  This implies that the parent matrix element is that for single gluon emission (which then undergoes a collinear splitting), whose spin-polarization tensor is
\begin{align}
T_g^{\mu \nu} (p) &= (4\pi\as\mu^{2\e}) \frac{4C_A}{p_T^2} \cdot \frac14 \bigl[ e^{2y_t} n^\mu n^\nu + e^{-2y_t} \bn^\mu \bn^\nu - (n^\mu \bn^\nu + n^\nu \bn^\mu) \bigr] \nn \\
&= T_g^{(0)} (p) \cdot \ca{E}^{\mu \nu} \,.
\end{align}
For a $q\bq$ initiated process, the color factor $C_A$ is traded for $C_F$.  Note that $(-g_{\mu \nu})\ca{E}^{\mu \nu} = 1$.  We will contract this spin-polarization tensor with the $\ord{\as^2}$ $g\to gg$, $g\to q\bq$ and the $\ord{\as^3}$ $g\to ggg$, $g\to gq\bq$ splitting functions.

%%%%%%%%%%%%%%%%%%%%%%%%%%%%%%%%%%%%%%%%%%%%%%%%%%%%%%%%%%%%%%%%%%%%%%%%%%%%%%%%%%%%%%%%%%%%%%%%%%%%%%%%%
\subsection{Splitting Functions}
\label{subsec:splittingfunctions}
%%%%%%%%%%%%%%%%%%%%%%%%%%%%%%%%%%%%%%%%%%%%%%%%%%%%%%%%%%%%%%%%%%%%%%%%%%%%%%%%%%%%%%%%%%%%%%%%%%%%%%%%%

The $1\to2$ and $1\to3$ polarization-dependent splitting functions are given in Refs.~\cite{Catani:1998nv,Catani:1999ss}.  One can also perform the average over polarizations \cite{Campbell:1997hg}.  For gluons,
\be
\langle \hat{P}_{a_1 \ldots a_n} \rangle = \ca{D}_{\mu\nu} \hat{P}_{a_1 \ldots a_n}^{\mu\nu} \,, \qquad \ca{D}_{\mu\nu} = \frac{1}{2(1-\e)} \biggl( - g_{\mu\nu} + \frac{\bn_p^\mu p^\nu + p^\mu \bn_p^\nu}{\bn_p \cdot p} \biggr) \,.
\ee
The polarization tensor $\ca{D}_{\mu\nu}$ is valid for axial gauge with $\bn_p \cdot A = 0$.

In principle, the polarization dependent terms can make a nonzero contribution to $C_3^{(2)}$.  However, this is not the case; we will show that the polarization dependent terms depend on orientation angles of the collinear system relative to the rest of the event that $C_3^{(2)}$ is insensitive to, and average to zero upon integration over the total phase space\footnote{Because the phase space constraints depend on an angular scale $R$, and the polarization average vanishing only holds in the collinear limit, the polarization dependent terms scale like powers of $R$.  Since we are already working in the small $R$ limit to extract the $\ln^2 R$ coefficient, we can safely neglect these contributions.}.  This is reasonable since the measurement functions that determine $C_3^{(2)}$ depend only on the relative positions of the collinear partons.  To see the polarization dependence explicitly, we will find the difference to the polarization averaged splitting functions, defining
\be
\hat{\ca{P}}_{a_1 \ldots a_n} \equiv \hat{P}_{a_1 \ldots a_n}^{\mu \nu} \ca{E}_{\mu \nu} = \langle \hat{P}_{a_1 \ldots a_n} \rangle + \Delta \hat{\ca{P}}_{a_1 \ldots a_n} \,, \qquad  \Delta \hat{\ca{P}}_{a_1 \ldots a_n} \equiv \hat{P}_{a_1 \ldots a_n}^{\mu \nu} \bigl( \ca{E}_{\mu \nu} - \ca{D}_{\mu \nu} \bigr) \,.
\ee
In \app{splittingfunctions} we give the relevant $\hat{P}_{a_1 \ldots a_n}^{ss'}$ and $\langle \hat{P}_{a_1 \ldots a_n} \rangle$, which originally appeared in Ref.~\cite{Catani:1998nv,Catani:1999ss}, and study $\Delta \hat{\ca{P}}_{a_1 \ldots a_n}$ here.

%%%%%%%%%%%%%
\subsubsection{$1\to2$ Splitting Functions and the Born Matrix Elements}
\label{subsec:1to2}
%%%%%%%%%%%%%

The $1\to2$ gluon-initiated splitting functions are given in \eq{1to2splitting}.  The polarization-dependent terms are
\begin{align}
g\to gg &: \quad \Delta \hat{\ca{P}}_{gg} = 2C_A \biggl[ z(1-z) \frac{\Delta y^2 - \Delta \phi^2}{\Delta y^2 + \Delta \phi^2} \biggr] \,, \nn \\
g\to q\bq &: \quad \Delta \hat{\ca{P}}_{q\bq} = T_F \biggl[ -2z(1-z) \frac{\Delta y^2 - \Delta \phi^2}{\Delta y^2 + \Delta \phi^2} \biggr] \,.
\end{align}
If $p_1$ and $p_2$ are the collinear momenta, then $z = x_1 / (x_1 + x_2)$ and $\Delta y, \Delta \phi$ are the rapidity and azimuthal angle separations.

The separations $\Delta y$ and $\Delta \phi$ make up the angle $\Delta R$, and we can define an angle $\theta$ which expresses how the collinear system is oriented relative to the transverse plane,
\be
\Delta y = \sin\theta \, \Delta R \,, \qquad \Delta \phi = \cos\theta \, \Delta R \,.
\ee
The measurement function and the polarization-averaged matrix elements are independent of $\theta$, which means that its only dependence is in the polarization dependent term $\Delta \hat{\ca{P}}$.  In the collinear limit, the integration over the full phase space will average over this angle and the polarization dependent terms will vanish:
\be
\int \df\Phi \, \Delta \hat{\ca{P}} \, \Delta \cM \varpropto \int_0^{2\pi} \frac{\df\theta}{2\pi} \cos 2\theta = 0 \,.
\ee
This justifies our use of the polarization-averaged $\ord{\as^2}$ splitting functions.  Hence the total Born matrix elements are (in 4 dimensions)
\begin{align} \label{eq:BornME}
\cT_{gg}^{(0)} (\Phi_2) &= (4\pi\as\mu^{2\e})^2 \, \frac{8 C_A}{p_{Tt}^2 s_{12}} \, \langle \hat{P}_{gg} \rangle \,, \nn \\
\cT_{q\bq}^{(0)} (\Phi_2) &= (4\pi\as\mu^{2\e})^2 \, \frac{8 C_A n_f}{p_{Tt}^2 s_{12}} \, \langle \hat{P}_{q\bq} \rangle \,.
\end{align}
If we include the symmetry factor of $\Sym_{gg} = 1/2!$ in the $gg$ matrix element, these agree with the known $\ord{\as^2}$ soft function matrix elements when taken into the collinear limit \cite{Hornig:2011iu}.  These can be used to calculate the $\ord{\as^2}$ clustering logarithm coefficient $C_2^{(1)}$, as shown in \app{C2}.

%%%%%%%%%%%%%
\subsubsection{$1\to3$ Splitting Functions and the Real Matrix Elements}
\label{subsec:1to3}
%%%%%%%%%%%%%

The story above repeats itself with the $\ord{\as^3}$ splitting functions.  The $g\to ggg$ and $g\to gq\bq$ splitting functions [given along with the polarization averages in \eq{1to3splitting}] have polarization dependent terms that depend on the orientation of the collinear system and average to zero upon integration over the phase space.  These polarization dependent terms, $\Delta \hat{\ca{P}}$, have no contribution from terms in $\hat{P}^{\mu\nu}$ proportional to $g^{\mu\nu}$, since $g^{\mu\nu} (\ca{E}_{\mu\nu} - \ca{D}_{\mu\nu}) = 0$.  All other terms have 3 basic structures, and we list their projection with $\ca{E}_{\mu\nu} - \ca{D}_{\mu\nu}$:
\begin{align}
k_{\perp 1}^\mu k_{\perp 1}^\nu (\ca{E}_{\mu\nu} - \ca{D}_{\mu\nu}) &= \frac12 z_1 (1-z_1) \bigl[ s_{12} \cos 2\theta_{12} + s_{13} \cos 2\theta_{13} + s_{23} \cos 2\theta_{23} \bigr] \nn \\
& \qquad - \frac12 z_1 s_{23} \cos 2\theta_{23} \,, \nn \\
k_{\perp 1}^\mu k_{\perp 2}^\nu (\ca{E}_{\mu\nu} - \ca{D}_{\mu\nu}) &= \frac14 \bigl[ z_1 s_{23} \cos 2\theta_{23} + z_2 s_{13} \cos 2\theta_{13} - z_3 s_{12} \cos 2\theta_{12} \bigr] \nn \\
& \qquad - \frac12 z_1 z_2 \bigl[ s_{12} \cos 2\theta_{12} + s_{13} \cos 2\theta_{13} + s_{23} \cos 2\theta_{23} \bigr] \,, \nn \\
z_1 z_2 \biggl( \frac{k_{\perp 1}}{z_1} - \frac{k_{\perp 2}}{z_2} \biggr)^\mu \biggl( \frac{k_{\perp 1}}{z_1} - \frac{k_{\perp 2}}{z_2} \biggr)^\nu &= \frac12 s_{12} \cos 2\theta_{12} \,.
\end{align}
Other choices of the indices on $k_{\perp}$ have the same form.  Each term contributing to $\Delta \hat{\ca{P}}$ is proportional to one of these structures, and each structure is in turn proportional to $\cos 2\theta_{ij}$ for some $i,j$.  This implies that, like at $\ord{\as}^2$, the polarization dependent terms cancel in the contribution to $C_3^{(2)}$.

The matrix elements for real emission are thus given in terms of the polarization averaged splitting functions,
\begin{align}
\cT^{(0)}_{ggg} &= (4\pi\as\mu^{2\e})^3 \frac{16 C_A}{p_{Tt}^2} \frac{1}{s_{123}^2} \langle \hat{P}_{ggg} \rangle \,, \nn \\
\cT^{(0)}_{gq\bq} &= (4\pi\as\mu^{2\e})^3 \frac{16 C_A n_f}{p_{Tt}^2} \frac{1}{s_{123}^2} \langle \hat{P}_{gq\bq} \rangle \,.
\end{align}
Note that there is no kinematic hierarchy used within the splitting functions (such as a strongly ordered limit), and that the complete $1\to3$ splitting functions are used in the matrix elements.

%%%%%%%%%%%%%%%%%%%%%%%%%%%%%%%%%%%%%%%%%%%%%%%%%%%%%%%%%%%%%%%%%%%%%%%%%%%%%%%%%%%%%%%%%%%%%%%%%%%%%%%%%
\section{Calculating Clustering Logarithms with Subtractions}
\label{sec:subtractions}
%%%%%%%%%%%%%%%%%%%%%%%%%%%%%%%%%%%%%%%%%%%%%%%%%%%%%%%%%%%%%%%%%%%%%%%%%%%%%%%%%%%%%%%%%%%%%%%%%%%%%%%%%

In the NLO calculation that determines $C_3^{(2)}$, there are canceling divergences in the real and virtual matrix elements, and the phase space cuts implemented by the measurement function are complex.  This situation is well-suited to use a subtraction to separately regulate the real and virtual contributions.  The universal soft and collinear factorization properties we have exploited to determine the matrix elements are precisely those which allow for subtractions that can regulate the real and virtual divergences simultaneously.

The basic form of a NLO calculation is
\begin{align}
\sigma_{\rm NLO} = \int \df\Phi_2 \, V_2 (\Phi_2) \cM^{(2)} (\Phi_2) + \int \df\Phi_3 \, B_3 (\Phi_3) \cM^{(3)} (\Phi_3) \,,
\end{align}
where $V_2$ and $B_3$ are the virtual and real matrix elements.  The soft, collinear, and collinear-soft divergences in the real emission cancel with the virtual contribution.  If we can define a set of subtraction terms $S_i (\Phi_3)$ that match the singularities of the real matrix element, then they can regulate both divergences at once:
\begin{align}
\sigma_{\rm NLO} &= \int \df\Phi_2 \bigl[ V_2 (\Phi_2) + I_S (\Phi_2) \bigr] \cM^{(2)} (\Phi_2)  \nn \\
& \qquad + \int \df\Phi_3 \bigl[ B_3 (\Phi_3) \cM^{(3)} (\Phi_3) - \sum_i S_i (\Phi_3) \cM^{(2)} (\Phi^i_2(\Phi_3)) \bigr] \,,
\end{align}
where
\be
I_S (\Phi_2) = \sum_i \int \frac{\df\Phi_3}{\df\Phi_2} S_i (\Phi_3) \,.
\ee
Note that the subtraction terms always come with the Born measurement function; this is because any infrared safe measurement cannot resolve a soft or collinear splitting in the singular limit (and hence depends on one fewer degree of freedom).  These Born events may differ for different singular limits.

Typically, the projection from $\Phi_3$ onto a $\Phi_2^i$ occurs via a map, using a phase space factorization formula to write $\Phi_3$ in terms of $\Phi_2^i$ and radiation variables.  In our case, however, the parent gluon in the splitting can have arbitrary momentum, meaning that the phase space naturally factorizes into radiation variables for each emission (basically making the map trivial).

The subtractions we use are essentially a variation of FKS subtractions, which separately handle soft, collinear, and collinear-soft singularities \cite{Frixione:1995ms,Frixione:2007vw}.  In FKS subtractions, these singularities are isolated into regions where at most one collinear and one soft singularity are present \cite{Frixione:1995ms}.  This is done by introducing the factor
\be
\cS_{ij} = \frac{1/s_{ij}}{\sum_{k,l} 1 / s_{kl}} \, h_{ij} \,,
\ee
with
\begin{align}
h_{ij} = \frac{E_j}{E_i + E_j} \,.
\end{align}
$\cS_{ij}$ is nonvanishing only if partons $i$ and $j$ become collinear or $i$ is soft (or the simultaneous soft-collinear limit).  The $\cS_{ij}$ satisfy $\sum_{i,j} \cS_{ij} = 1$ and the following limits:
\begin{align} \label{eq:sectorlimits}
\cS_{ij} \to \begin{cases}
\; \cS_{ij}^{\rm coll} = h_{ij} & \textrm{ if } i \textrm{ and } j \textrm{ become collinear, } \\  \vspace{0.5ex}
\cS_{ij}^{i \text{ soft }} = \dfrac{1/s_{ij}}{\sum_k 1/s_{ik}} & \textrm{ if } i \textrm{ becomes soft, } \\
\cS_{ij}^{\rm cs} = 1 & \textrm{ if } i \textrm{ becomes soft, and } i,j \textrm{ become collinear, } \\
\;0 & \textrm{ if } j \textrm{ becomes soft, } \\
\;0 & \textrm{ if } s_{kl} \to 0 \textrm{ for } i,j \neq k,l.
\end{cases}
\end{align}
This machinery is useful once we pair it with the subtraction terms.  In our case, we will be able to write the singular limits of the matrix element in terms of factors that multiply the Born matrix element.  These subtraction factors are:
\begin{align}
S_{kl}^i &: \; \textrm{soft radiation of particle $i$ between particles $k$ and $l$} \,, \nn \\
C_{ij} &: \; \textrm{collinear splitting into particles $i$ and $j$ $(i < j)$} \,, \nn \\
\CS_{ij} &: \; \textrm{soft-collinear splitting into particles $i$ and $j$, with $i$ soft} \,. \nn 
\end{align}
And the real matrix element becomes, under specific singular limits,
\begin{align}
\cT^{(0)}_{a_1 a_2 a_3} (k_1, k_2, k_3) \; \longrightarrow \; \begin{cases}
\textrm{ 3 soft: } & \cT^{(0)}_{a_1 a_2} (k_1, k_2) \sum_{k,l} S_{12}^3 (k_1, k_2, k_3) \,, \\
\textrm{ 1,3 collinear: } & \cT^{(0)}_{a_{13} a_2} (k_1 + k_3, k_2) \, C_{13} (k_1, k_3) \,, \\
\textrm{ 3 soft + 1,3 collinear: } & \cT^{(0)}_{a_1 a_2} (k_1, k_2) \, \CS_{31} (k_1, k_3) \,.
\end{cases}
\end{align}
The above assumes that $a_3 = g$, so that soft and collinear-soft limits are actually singular.  We can take the singular limits of the real matrix elements analytically and define the subtractions in terms of plus distributions.  We will explicitly use the $g\to ggg$ matrix elements as an example in the rest of this section, and the general case easily follows.  Integrating the subtractions is performed in \app{integratedsubtractions}.

Each subtraction will regulate a particular divergence in the real matrix element, and the integrated subtraction compensates for the terms that are introduced.  It is often very useful to introduce auxiliary cuts on the singular variables in the subtractions and integrated subtractions, which serve as a self-consistency check on the calculation (the dependence on these cuts should cancel in the total cross section) and can help probe the singularity structure (in our case, the dependence on $R$).  In the usual FKS subtractions, these cuts are on the soft gluon energy in soft and collinear-soft subtractions and the splitting angle in the collinear and collinear-soft subtractions.  For example, instead of integrating over all soft gluon energies $E_g$, one integrates over the range $0 < E_g < E_c$.  The dependence on the artificial parameter $E_c$ must cancel between the subtraction and integrated subtraction.  In our case, we will find the following cuts useful:
\begin{align}
x_g < x_c \; &: \; \textrm{ for soft subtractions, with } x_g \textrm{ the soft gluon } p_T \textrm{ fraction} \,, \nn \\
\Delta R < R_c \; &: \; \textrm{ for collinear subtractions, with } \Delta R \textrm{ the opening angle of the splitting} \,.
\end{align}
The collinear-soft subtraction will have both of these cuts, and the integrated subtractions will depend on $x_c$ and $R_c$ in the appropriate places.

The matrix elements for the subtractions are closely related to matrix elements in SCET.  This is not surprising, as both are built from the singular limits of QCD: we show that the soft, collinear, and collinear-soft FKS subtractions are given by soft, naive jet, and zero-bin jet function matrix elements.  We will explore these connections further in this section.

%%%%%%%%%%%%%%%%%%%%%%%%%%%%%%%%%%%%%%%%%%%%%%%%%%%%%%%%%%%%%%%%%%%%%%%%%%%%%%%%%%%%%%%%%%%%%%%%%%%%%%%%%
\subsection{Soft Subtractions}
\label{subsec:softsubtractions}
%%%%%%%%%%%%%%%%%%%%%%%%%%%%%%%%%%%%%%%%%%%%%%%%%%%%%%%%%%%%%%%%%%%%%%%%%%%%%%%%%%%%%%%%%%%%%%%%%%%%%%%%%

In the limit that gluon 3 becomes soft, the $g\to ggg$ matrix element becomes
\begin{align} \label{eq:Tgggsoftlimit}
\cT_{ggg}^{(0)} (3 \to \textrm{ soft}) \to (4\pi\as \mu^{2\e}) \frac{2C_A}{p_{T3}^2} \biggl( \frac{\Delta R_{12}^2}{\Delta R_{13}^2 \Delta R_{23}^2} + \frac{1}{\Delta R_{13}^2} + \frac{1}{\Delta R_{23}^2} \biggr) \, \cT_{gg}^{(0)} (k_1, k_2) \,.
\end{align}
This expression can be understood as a sum of eikonal factors multiplied by the Born matrix element for gluons 1 and 2.  The first term is the eikonal factor for soft gluon exchange between gluons 1 and 2 (proportional to the Born matrix element), which defines the subtraction term,
\begin{align} \label{eq:S12}
S_{12}^3 + S_{21}^3 = (4\pi\alpha_s \mu^{2\e}) \frac{2C_A}{p_{T3}^2} \frac{\Delta R_{12}^2}{\Delta R_{13}^2 \Delta R_{23}^2} = (4\pi\as \mu^{2\e}) (-2\T_1\cdot\T_2) \frac{k_1 \cdot k_2}{(k_1 \cdot k_3)(k_2 \cdot k_3)} \,.
\end{align}
For $g\to ggg$, the color operators $\T_1$ and $\T_2$ obey the relations $\T_1^2 = \T_2^2 = C_A$ and $(\T_1 + \T_2)^2 \equiv \T_t^2 = C_A$, where $\T_t$ is the color operator for the initial gluon that splits.  It is also useful to define the color operator $\T_r \equiv -\T_t$ for the rest of the event, which also obeys $\T_r^2 = C_A$.  The subtraction term in \eq{S12} can be recognized as the $\ord{\as}$ soft function matrix element for soft gluon exchange between soft Wilson lines 1 and 2.  This is natural, as both the subtraction and the soft function matrix elements exploit the eikonal factorization properties of QCD amplitudes.

The second and third terms in \eq{Tgggsoftlimit} come from soft gluon exchange between either gluon 1 or gluon 2 and the rest of the event.  Because the $g\to ggg$ system is color-connected to the rest of the event, other partons can radiate soft gluons into the collinear system.  If there is another parton $i$ that exchanges soft gluon 3 with gluon 1, then the usual eikonal factor is
\be
(4\pi\as \mu^{2\e}) (-2\T_1 \cdot \T_i) \frac{k_i \cdot k_1}{(k_i \cdot k_3)(k_1 \cdot k_3)} \,.
\ee
However, since gluon 3 must be collinear with gluon 1, $\theta_{13} \ll \theta_{i3}$ and
\be
\frac{k_i\cdot k_1}{k_i\cdot k_3} = \frac{E_1}{E_3} + \ord{\theta_{13}^2} \,,
\ee
and thus the eikonal factor reduces to (using the collinear limit)
\begin{align}
(4\pi\as \mu^{2\e}) (-2\T_1 \cdot \T_i) \frac{k_i \cdot k_1}{(k_i \cdot k_3)(k_1 \cdot k_3)} &= (4\pi\as \mu^{2\e}) (-2\T_1 \cdot \T_i) \frac{E_1}{E_3} \frac{1}{k_1 \cdot k_3} \nn \\
&= (4\pi\as \mu^{2\e}) (-2\T_1 \cdot \T_i) \frac{2}{p_{T3}^2 \Delta R_{13}^2} \,.
\end{align}
The kinematic factor is $i$-independent, and so we can sum over all colors.  This is an example of coherent soft gluon emission by the rest of the event, and is well-studied in related contexts \cite{Ellis:1991qj}.  This yields
\be \label{eq:S1r}
S_{1r}^3 + S_{r1}^3 = (4\pi\as \mu^{2\e}) (-2\T_1 \cdot \T_r) \frac{2}{p_{T3}^2 \Delta R_{13}^2} = (4\pi\as \mu^{2\e}) (-2\T_1 \cdot \T_r) \frac{1}{(k_3^0)^2}\frac{1}{(n_1 \cdot n_3)} \,,
\ee
where $-2\T_1 \cdot \T_r = C_A$ for $g\to ggg$, which matches the term in \eq{Tgggsoftlimit}.  

Note that this subtraction term is actually in a collinear-soft limit.  This is arising in the soft subtraction because we are demanding the final state partons are collimated, necessitating the additional expansion.  Thus, these subtractions will reappear in the collinear-soft case (where they remove double counting of divergences with the collinear subtraction), and in that case the color connections are different (so we will retain distinct labels for $S^3_{1r}$ and $\CS$).  The matrix element in \eq{S1r} can be recognized as the zero-bin matrix element in the jet function, which coincides with the soft function matrix element taken into the collinear limit (where the soft gluon is collinear to one of the Wilson lines).

This combined collinear-soft limit has also been studied previously in SCET \cite{Bauer:2011uc}, and can be understood in terms of an additional factorization in precisely the collinear limit that we are studying.  That is, soft gluons in the splitting function also have collinear scaling, and Ref.~\cite{Bauer:2011uc} terms them \emph{csoft} gluons.  Since the emission of these csoft gluons is controlled by the collinear system, as well as coherent soft radiation from the rest of the event, their emission factorizes at the level of operators in SCET using fundamentally the same collinear factorization properties used to write the overall matrix elements.  The total soft gluon emission matrix element, in \eq{Tgggsoftlimit}, is given precisely by the csoft function as formulated in Ref.~\cite{Bauer:2011uc}.  In the csoft function the ``rest of the event'' is represented by a soft Wilson line $(V)$ that radiates soft gluons from the anti-collinear direction.

%%%%%%%%%%%%%%%%%%%%%%%%%%%%%%%%%%%%%%%%%%%%%%%%%%%%%%%%%%%%%%%%%%%%%%%%%%%%%%%%%%%%%%%%%%%%%%%%%%%%%%%%%
\subsection{Collinear Subtractions}
\label{subsec:collinearsubtractions}
%%%%%%%%%%%%%%%%%%%%%%%%%%%%%%%%%%%%%%%%%%%%%%%%%%%%%%%%%%%%%%%%%%%%%%%%%%%%%%%%%%%%%%%%%%%%%%%%%%%%%%%%%

In the collinear limit of gluons 1 and 3, the matrix element takes the form
\begin{align}
\cT_{ggg}^{(0)} (1,3 \to \textrm{ collinear}) \to (4\pi\as\mu^{2\e}) \frac{2}{s_{13}} \langle \hat{P}_{gg} (1,3) \rangle \cT_{gg}^{(0)} (k_1 + k_3, k_2) + \textrm{ azimuthal terms} \,,
\end{align}
where the azimuthal terms are dependent on the polarization of the 1-3 collinear system relative to gluon 2 \cite{Knowles:1988vs}, and vanish in the total contribution to $C_3^{(2)}$.  The polarization-independent terms are given by tree-level collinear factorization applied to the 1-3 leg of the underlying Born matrix element $\cT_{gg}^{(0)} (k_1 + k_3, k_2)$, and the subtraction factor is
\be \label{eq:C13}
C_{13} (k_1, k_3) = (4\pi\as\mu^{2\e}) \frac{2}{s_{13}} \langle \hat{P}_{gg} (1,3) \rangle \,.
\ee
More generally, we need to choose the appropriate splitting function for the process,
\be \label{eq:Cij}
C_{ij} (k_i, k_j) = (4\pi\as\mu^{2\e}) \frac{2}{s_{ij}} \langle \hat{P}_{ij} (i,j) \rangle \,.
\ee
This subtraction term may be recognized as the $\ord{\as}$ jet function matrix elements (before a zero-bin subtraction \cite{Manohar:2006nz}).  As with the soft subtraction, this is expected as the jet function is based on the same collinear factorization properties (see Ref.~\cite{Feige:2013zla}) as the collinear subtraction terms.

%%%%%%%%%%%%%%%%%%%%%%%%%%%%%%%%%%%%%%%%%%%%%%%%%%%%%%%%%%%%%%%%%%%%%%%%%%%%%%%%%%%%%%%%%%%%%%%%%%%%%%%%%
\subsection{Collinear-Soft Subtractions}
\label{subsec:collinearsoftsubtractions}
%%%%%%%%%%%%%%%%%%%%%%%%%%%%%%%%%%%%%%%%%%%%%%%%%%%%%%%%%%%%%%%%%%%%%%%%%%%%%%%%%%%%%%%%%%%%%%%%%%%%%%%%%

The collinear-soft limit can be taken either from the soft or collinear limits above, and there are particular soft subtractions (those where the soft gluon is exchanged with the ``rest of the event'') whose kinematic dependence matches the collinear-soft subtraction (see \subsec{softsubtractions}).  The collinear-soft limit accounts for the double counting of divergences between collinear and soft, and is
\begin{align} \label{eq:Tgggcsoftlimit}
\cT_{ggg}^{(0)} (3 \to \textrm{ soft}, \, 1,3 \textrm{ collinear}) &\to (4\pi\as \mu^{2\e}) \frac{4C_A}{p_{T3}^2} \biggl( \frac{1}{\Delta R_{13}^2} \biggr) \, \cT_{gg}^{(0)} (k_1 + k_3, k_2) \nn \\
& \quad = (4\pi\as\mu^{2\e}) \frac{2}{s_{13}} \langle \hat{P}^{(0)}_{gg} (1,3) \rangle \cT_{gg}^{(0)} (k_1 + k_3, k_2) \,.
\end{align}
Above $P^{(0)}_{gg}$ is the soft limit of the $g\to gg$ splitting function.  The subtraction factor is, in terms of splitting functions,
\be
\CS_{31} (k_1, k_3) = (4\pi\as\mu^{2\e}) \frac{2}{s_{13}} \langle \hat{P}^{(0)}_{gg} (1,3) \rangle \,,
\ee
and more generally the $g\to gg$ splitting function is replaced by the right one for the process.  It can also be written in the form of \eq{S1r} (with a different color factor),
\be \label{eq:CSsub}
\CS_{31} (k_1, k_3) = (4\pi\as\mu^{2\e}) \, 2\T_1^2 \frac{1}{(k_3^0)^2} \frac{1}{(n_1 \cdot n_3)} \,.
\ee
As discussed in \subsec{softsubtractions}, this subtraction term is given by the zero-bin matrix element in the jet function, or equivalently by the $\ord{\as}$ soft function taken into the limit where the soft gluon is collinear to one of the Wilson lines.

%%%%%%%%%%%%%%%%%%%%%%%%%%%%%%%%%%%%%%%%%%%%%%%%%%%%%%%%%%%%%%%%%%%%%%%%%%%%%%%%%%%%%%%%%%%%%%%%%%%%%%%%%
\subsection{The Regulated Real Emission}
\label{subsec:regreal}
%%%%%%%%%%%%%%%%%%%%%%%%%%%%%%%%%%%%%%%%%%%%%%%%%%%%%%%%%%%%%%%%%%%%%%%%%%%%%%%%%%%%%%%%%%%%%%%%%%%%%%%%%

Having detailed the subtractions, we can now use them to regulate the real matrix elements.  The sum over all regions is
\be
\Sym_{a_1 a_2 a_3} \cT_{a_1 a_2 a_3}^{(0)} = (\cS_{12} + \cS_{21} + \cS_{13} + \cS_{31} + \cS_{23} + \cS_{32}) \, \Sym_{a_1 a_2 a_3} \cT_{a_1 a_2 a_3}^{(0)} \,.
\ee
For the $g\to ggg$ case, the exchange symmetry means that we can drop the symmetry factor by choosing one region,
\be
(\cS_{12} + \cS_{21} + \cS_{13} + \cS_{31} + \cS_{23} + \cS_{32}) \, \Sym_{a_1 a_2 a_3} \cT_{ggg}^{(0)} \; \backsimeq \; \cS_{31} \cT_{ggg}^{(0)} \,.
\ee
This equivalence is obviously not point-by-point, but is true integrated over phase space.  This limits the number of subtractions we need, since $\cS_{31} \cT_{ggg}^{(0)}$ is singular only in a fraction of the limits that $\cT_{ggg}^{(0)}$ is.  Using the limits in \eq{sectorlimits}, the $ggg$ channel is regulated in the combination
\begin{align} \label{eq:Rggg}
R_{ggg} \, &: \;\cS_{31} \cT_{ggg}^{(0)} (k_1, k_2, k_3) - \Bigl\{ \cS_{31}^{\rm soft} \sum_{\substack{i\neq j \\ \{1,2,r\}}} S^3_{ij} \, \cT_{gg}^{(0)} (k_1, k_2) \nn \\
& \qquad \qquad + \cS_{31}^{\rm coll} C_{13} \, \cT_{gg}^{(0)} (k_1 + k_3, k_2) - \cS_{31}^{\rm cs} \CS_{31} \, \cT_{gg}^{(0)} (k_1, k_2) \Bigr\} \,,
\end{align}
and the contribution to the soft function we call $R_{ggg}$.  Note that the soft and collinear limits of the $\cS_{ij}$, $\cS_{31}^{\rm soft}$ and $\cS_{31}^{\rm coll}$, average to $1/2$ in integrating over the Born phase space.

For the $gq\bq$ case all the subtractions must be included (although only the gluon can become soft or collinear-soft).  For this channel we switch to the labels $g$, $q$, and $\bq$.  In each sector we take the limits for each subtraction, and the regulated combination is
\begin{align} \label{eq:Rgqq}
R_{gq\bq} \, &: \; \sum_{i\neq j} \cS_{ij} \cT_{gq\bq}^{(0)} (k_g, k_q, k_\bq) - \Bigl\{ \sum_{\substack{i\neq j \\ \{q,\bq,r\}}} S^g_{ij} \, \cT_{q\bq}^{(0)} (k_q, k_\bq) + C_{gq} \, \cT_{q\bq}^{(0)} (k_g + k_q, k_\bq) \\
& \qquad + C_{g\bq} \, \cT_{q\bq}^{(0)} (k_g + k_\bq, k_q) + C_{q\bq} \, \cT_{gg}^{(0)} (k_g, k_q + k_\bq) - \bigl[\CS_{gq} + \CS_{g\bq} \bigr] \, \cT_{q\bq}^{(0)} (k_q, k_\bq) \Bigr\} \,. \nn
\end{align}
We find that $R_{ggg}$ and $R_{gq\bq}$ are indeed finite when integrated over phase space.  These regulated pieces will be calculated numerically, and the calculation discussed in \sec{calc}.

%%%%%%%%%%%%%%%%%%%%%%%%%%%%%%%%%%%%%%%%%%%%%%%%%%%%%%%%%%%%%%%%%%%%%%%%%%%%%%%%%%%%%%%%%%%%%%%%%%%%%%%%%
\section{Virtual Matrix Elements and Integrated Subtractions}
\label{sec:virt}
%%%%%%%%%%%%%%%%%%%%%%%%%%%%%%%%%%%%%%%%%%%%%%%%%%%%%%%%%%%%%%%%%%%%%%%%%%%%%%%%%%%%%%%%%%%%%%%%%%%%%%%%%

So far, we have made use of several factorization properties to define the real matrix elements in terms of splitting functions.  Extracting the corresponding virtual contribution is more involved.  We are not calculating a ``complete'' cross section, in the sense that we are neglecting subleading terms in $\ln R$, and hence only including matrix elements singular in the appropriate limit.  Our approach is to focus only on the terms which give a contribution to $C_3^{(2)}$, which are directly related to one-loop splitting amplitudes.

In this section we will show that the only terms in the virtual matrix elements that can make a contribution to $C_3^{(2)}$ are particular terms that come from the one-loop splitting amplitudes.  Pairing their dependence with the integrated subtractions we can determine the contribution to $C_3^{(2)}$ analytically.

%%%%%%%%%%%%%%%%%%%%%%%%%%%%%%%%%%%%%%%%%%%%%%%%%%%%%%%%%%%%%%%%%%%%%%%%%%%%%%%%%%%%%%%%%%%%%%%%%%%%%%%%%
\subsection{The Virtual Matrix Elements}
\label{subsec:virtMEs}
%%%%%%%%%%%%%%%%%%%%%%%%%%%%%%%%%%%%%%%%%%%%%%%%%%%%%%%%%%%%%%%%%%%%%%%%%%%%%%%%%%%%%%%%%%%%%%%%%%%%%%%%%

The collinear factorization property in \eq{collfact} extends beyond tree level, although more naturally in terms of amplitudes.  We use the notation in Ref.~\cite{Catani:1998nv,Catani:1999ss}.  If $\ket{\cM^{(0)}_{n+1} (p_1, p_2, \ldots)}$ is the $n+1$-particle tree-level amplitude which is a vector in color and spin space, then in the collinear limit of particles $p_1$ and $p_2$ the factorization is
\be \label{eq:treefact}
\ket{\cM^{(0)}_{n+1} (p_1, p_2, \ldots)} \; \xrightarrow[1,2 \textrm{ coll.}]{} \; \Sp^{(0)} (p_1, p_2, P) \ket{\cM^{(0)}_n (P, \ldots)} \,,
\ee
where $P$ is the collinear limit of $p_1 + p_2$.  The splitting matrix $\Sp^{(0)}$ is a matrix in color and spin space, and is related to the tree-level splitting amplitude $\Split^\tree$ via a simple color matrix.  At one loop, the virtual corrections satisfy a similar property (see, e.g., Ref.~\cite{Bern:1994zx}):
\begin{align} \label{eq:1loopfact}
\ket{\cM^{(1)}_{n+1} (p_1, p_2, \ldots)} \; \xrightarrow[1,2 \textrm{ coll.}]{} &\;  \Sp^{(1)} (p_1, p_2, P) \ket{\cM^{(0)}_n (P, \ldots)} \\
& \qquad\qquad + \Sp^{(0)} (p_1, p_2, P) \ket{\cM^{(1)}_n (P, \ldots)} \,. \nn
\end{align}
That is, the loop corrections to the unfactorized amplitude divide into loop corrections to the splitting matrix and loop corrections to the underlying $n$-particle amplitude.  This factorization is divided into two terms, and the crucial property in our case will be that the $s_{12}$ dependence is isolated into the tree-level and 1-loop splitting matrices.

Furthermore, the 1-loop amplitudes for the $n$-parton configuration (where the collinear partons have been clustered) satisfy the decomposition
\be
\ket{\cM^{(1)}_n} = I_n^{(1)}\ket{\cM^{(0)}_n} + \ket{\cM^{(1)\, \rm fin}_n} \,,
\ee
where $I_n^{(1)}$ is a color operator containing the IR poles and $\ket{\cM^{(1)\, \rm fin}_n}$ is finite as $\e\to0$.  The same decomposition holds for the 1-loop splitting amplitude,
\be
\Sp^{(1)} = I_C^{(1)} \, \Sp^{(0)} + \Sp_H^{(1)} \,,
\ee
where $I_C^{(1)}$ contains the IR poles and $\Sp_H^{(1)}$ is finite as $\e\to0$, containing only rational dependence on the splitting momenta.

Putting these forms into the 1-loop collinear factorization formula, we have
\begin{align}
\ket{\cM^{(1)}_{n+1} (p_1, p_2, \ldots)} \; \xrightarrow[1,2 \textrm{ coll.}]{} &\;  \bigl( I_C^{(1)} + I_n^{(1)} \bigr) \Sp^{(0)} (p_1, p_2, P) \ket{\cM^{(0)}_n (P, \ldots)} \\
& \qquad\qquad + \Sp^{(0)} (p_1, p_2, P) \ket{\cM^{(1)\, \rm fin}_n (P, \ldots)} \nn \\
& \qquad\qquad + \Sp_H^{(1)} (p_1, p_2, P) \ket{\cM^{(0)}_n (P, \ldots)} \,. \nn
\end{align}
The last two lines are finite as $\e\to0$ and the only singular dependence on $s_{12}$ comes from the $1/s_{12}$ present in the splitting matrices.  It will become evident below that we can neglect these terms.  The first line is a reflection of the fact that the IR poles are proportional to the Born amplitude.  Indeed, the matrix element squared is
\begin{align} \label{eq:T1loop}
\cT_{n+1}^{(1)} (\Phi_{n+1}) &= \sum_{\substack{\rm spins, \\ \rm colors}} \Bigl( \braket{\cM^{(1)}_{n+1} (p_1, p_2, \ldots)}{\cM^{(0)}_{n+1} (p_1, p_2, \ldots)} \nn \\
& \qquad \qquad + \braket{\cM^{(0)}_{n+1} (p_1, p_2, \ldots)}{\cM^{(1)}_{n+1} (p_1, p_2, \ldots)} \Bigr) + \textrm{ finite} \nn \\
& = \bigl( I_C^{(1)\, \dag} + I_C^{(1)} + I_n^{(1)\, \dag} + I_n^{(1)} \bigr) \bra{\cM^{(0)}_n (P, \ldots)} \bigl( \Sp^{(0)\, \dag} \Sp^{(0)} \bigr) (p_1, p_2, P) \ket{\cM^{(0)}_n (P, \ldots)}  \nn \\
& \qquad \qquad + \textrm{ finite} \nn \\
& = \bigl( I_C^{(1)\, \dag} + I_C^{(1)} + I_n^{(1)\, \dag} + I_n^{(1)} \bigr) \cT_n^{(0)} (\Phi_n) + \textrm{ finite} \,.
\end{align}
The integrated subtraction is also proportional to the Born matrix element, and the IR singularities cancel at the level of the prefactor to the Born matrix element.  That is, since the virtuals and integrated subtractions share the Born matrix element as a prefactor to the poles, including subleading terms in $\e$ (that vanish when $\e\to0$), we do not have to worry about $1/\e^2, 1/\e$ poles multiplying subleading terms in $\e$ in the Born matrix element to generate finite terms.  This is well known, but it is a crucial property that we must exploit in determining the contribution of the virtual matrix elements to $C_3^{(2)}$.

Let us consider what kinematic dependence in the virtual and integrated subtractions can give rise to $\ln^2 R$ terms.  Since both contributions are proportional to the Born matrix element, the relevant question is what kinematic dependence is required in the prefactor.  It is clear that the logarithms of $R$ arise from the angular phase space integrals; in the Born contribution the relevant integrals are
\be \label{eq:lnRint}
\int_{-\infty}^{\infty} \df y \int_{-\pi}^{\pi} \frac{\df\phi}{2\pi} \frac{1}{\Delta R^2} \theta(\Delta R > R) = - \ln \frac{R}{2\pi} \,.
\ee
The $1/\Delta R^2$ factor arises from the propagator of the collinear parton that splits, which gives a factor of $1/s$.  To get another logarithm of $R$, we must have a factor of $\ln \Delta R$, as
\be \label{eq:lnRsqint}
\int_{-\infty}^{\infty} \df y \int_{-\pi}^{\pi} \frac{\df\phi}{2\pi} \frac{1}{\Delta R^2} \ln \Delta R \, \theta(\Delta R > R) = -\frac12 \ln^2 R + \textrm{ constant} \,.
\ee
In the virtual matrix elements, the $\Delta R^2$ dependence arises only from the invariant mass $s_{12}$ of the collinear pair of partons.  Therefore, only finite terms in the virtual matrix elements that contain a logarithm of $s_{12}$ can contribute to $C_3^{(2)}$.  This means we only need to track this dependence in the virtuals, and do not need the complete one-loop matrix elements.  Since the splitting matrix $\Sp^{(1)}$, and in particle $I_C$, is the only part of the virtuals that depend on $s_{12}$, we can focus only on this term.

The divergent terms in the 1-loop splitting matrix are \cite{Kosower:1999rx}
\begin{align} \label{eq:IC}
I_C^{(1)} &= \frac{\as}{4\pi} \, c_\Gamma \biggl( \frac{\mu^2}{-s_{12} - \img0} \biggr)^{\e} \biggl\{ \frac{1}{\e^2} \bigl( C_{12} - C_1 - C_2 \bigr) + \frac{1}{\e} \bigl( \gamma_{12} - \gamma_1 - \gamma_2 + \frac12 \beta_0 \bigr) \nn \\
& \qquad \qquad - \frac{1}{\e} \Bigl[ \bigl( C_{12} + C_1 - C_2 \bigr) f(\e; z) + \bigl( C_{12} - C_1 + C_2 \bigr) f(\e; 1-z) \Bigr] \biggr\} \,,
\end{align}
where, in $\overline{\rm MS}$, $c_\Gamma = e^{\gamma_E \e} \Gamma(1+\e) \Gamma^2 (1-\e) / \Gamma(1-2\e) = 1 + \ord{\e^2}$ and 
\be 
f(\e; z) = - \ln z + \e \Bigl[ \frac12 \ln^2 z + \Li_2 (1 - z) \Bigr] + \ord{\e^2} \,,
\ee
with $z = x_1 / (x_1 + x_2)$.  The constants $C_i$ are the Casimirs for each parton in the splitting, with $C_g = C_A$ and $C_q = C_F$, and the constants $\gamma_i$ are
\be
\gamma_g = \frac{11}{6} C_A - \frac46 T_F n_f \,, \qquad\qquad \gamma_q = \frac32 C_F \,.
\ee
The $\beta_0 / 2$ single pole in \eq{IC} is removed if the virtual matrix elements are renormalized (as we do below), but a remnant finite term remains\footnote{See \emph{Note Added} at the end of \sec{calc}.}

For the $gg$ channel, $1 = 2 = 12 = g$ and so the relevant factor in the virtual matrix elements is, including the symmetry factor $\Sym_{gg} = 1/2!$ for the $gg$ channel,
\begin{align} \label{eq:IVgg}
\frac{1}{2!} \biggl( I_{C,gg}^{(1)\, \dag} + I_{C,gg}^{(1)} \biggr) = \frac{\as}{2\pi} \biggl(\frac{\mu}{\pTcut} \biggr)^{2\e} & \biggl\{ - \frac{1}{2\e^2} C_A + \frac{1}{2\e} \biggl[ C_A \ln \biggl(\frac{x_1^2 x_2^2}{(x_1 + x_2)^2} \Delta R_{12}^2 \biggr) - \gamma_g^{(g)} \biggr] \nn \\
& \qquad - \ln^2 \Delta R_{12} + \ln \Delta R_{12} \biggl[ \gamma_g - \frac12 \beta_0 - 2 \ln \frac{x_1 x_2}{x_1 + x_2} \biggr] \nn \\
& \qquad + \Delta R_{12}\textrm{-independent, finite} \biggr\} \,.
\end{align}
which multiplies the Born matrix element $\cT_{gg}^{(0)}$.  The integrated subtractions must cancel the $\Delta R_{12}$ dependence in the divergent terms as well as the finite $\ln^2 \Delta R_{12}$ term (which would generate a $\ln^3 R$).  The $q\bq$ channel has $1,2 = q,\bq$ and $12 = g$, meaning the relevant factor in the virtual matrix elements is
\begin{align} \label{eq:IVqq}
& I_{C,q\bq}^{(1)\, \dag} + I_{C,q\bq}^{(1)} = \frac{\as}{2\pi} \biggl(\frac{\mu}{\pTcut} \biggr)^{2\e} \biggl\{ -(2C_F - C_A) \frac{1}{\e^2} \nn \\
& \qquad + \frac{1}{\e} \biggl[ (2C_F - C_A) \ln x_1 x_2 \Delta R_{12}^2 + C_A \ln \frac{x_1 x_2}{(x_1 + x_2)^2} + (\gamma_g - 2\gamma_q) \biggr] - 2(2C_F - C_A) \ln^2 \Delta R_{12} \nn \\
& \qquad + \ln \Delta R_{12} \biggl[ -2(2C_F - C_A) \ln x_1 x_2 - 2C_A \ln \frac{x_1 x_2}{(x_1 + x_2)^2} - 2(\gamma_g - 2\gamma_q + \beta_0 / 2) \biggr] \nn \\
& \qquad  + \Delta R_{12}\textrm{-independent, finite} \biggr\} \,.
\end{align}

%%%%%%%%%%%%%%%%%%%%%%%%%%%%%%%%%%%%%%%%%%%%%%%%%%%%%%%%%%%%%%%%%%%%%%%%%%%%%%%%%%%%%%%%%%%%%%%%%%%%%%%%%
\subsection{Adding in the Integrated Subtractions}
\label{subsec:intsub}
%%%%%%%%%%%%%%%%%%%%%%%%%%%%%%%%%%%%%%%%%%%%%%%%%%%%%%%%%%%%%%%%%%%%%%%%%%%%%%%%%%%%%%%%%%%%%%%%%%%%%%%%%

Similar to the virtual corrections, the terms in the integrated subtractions that can contribute to $C_3^{(2)}$ are those with an additional $\ln \Delta R_{12}$ multiplying the Born matrix element.  The integration subtractions are found in \app{integratedsubtractions}, and it can be seen that only the soft subtractions with soft gluon exchange between particles 1 and 2 depend on $\Delta R_{12}$.  Since soft gluon emissions from the $gg$ and $q\bq$ Born configurations contribute to the $ggg$ and $gq\bq$ channels respectively, this means the integrated soft subtractions in the $ggg$ channel should be paired with the $gg$ virtual matrix elements, and the $gq\bq$ integrated subtractions paired with the $q\bq$ virtual matrix elements\footnote{If we were accounting for all divergences, we would have to take into account the fact that the $gq\bq$ channel can have divergences originating from the $gg$ channel via a $g\to q\bq$ splitting, although these do not contribute to $C_3^{(2)}$.}.

Using \eqs{Rggg}{Rgqq} to determine the relevant $\Delta R_{12}$ dependent terms, we find
\begin{align} \label{eq:Iggg}
I_{\rm tot}^{ggg} &\supset \frac{\as}{2\pi} \biggl( \frac{\mu}{\pTcut} \biggr)^{2\e} C_A \biggl\{ -\frac{1}{\e} \ln \Delta R_{12} + \ln^2 \Delta R_{12} + 2 \ln x_c \ln \Delta R_{12} \biggr\} \,,
\end{align}
which multiplies $\cT_{gg}^{(0)}$, and
\begin{align} \label{eq:Igqq}
I_{\rm tot}^{gq\bq} &\supset \frac{\as}{2\pi} \biggl( \frac{\mu}{\pTcut} \biggr)^{2\e} (2C_F - C_A) \biggl\{ - \frac{2}{\e} \ln \Delta R_{12} + 2 \ln^2 \Delta R_{12} + 4 \ln x_c \ln \Delta R_{12} \biggr\} \,,
\end{align}
which multiplies $\cT_{q\bq}^{(0)}$.  Adding these integrated subtractions to the virtual matrix elements in \eqs{IVgg}{IVqq}, we obtain the regulated virtual matrix elements $V^S_{gg}$ and $V^S_{q\bq}$ that contribute to $C_3^{(2)}$.  For the $gg$ channel, we have
\begin{align} \label{eq:VSgg}
V^S_{gg} (\Phi_2) &= \frac{\as}{2\pi} \ln \Delta R_{12} \biggl[ 2C_A \ln \biggl( \frac{x_1 + x_2}{x_1 x_2} \, x_c \biggr) + \gamma_g - \beta_0 / 2 \biggr] \cT_{gg}^{(0)} (\Phi_2) \,,
\end{align}
and for the $q\bq$ channel,
\begin{align} \label{eq:VSqq}
V^S_{q\bq} (\Phi_2) &= \frac{\as}{2\pi} \ln \Delta R_{12} \biggl[ 2(2C_F - C_A) \ln \frac{x_c^2}{x_1 x_2} - 2C_A \ln \frac{x_1 x_2}{(x_1 + x_2)^2} + 2(2\gamma_q - \gamma_g - \beta_0 / 2) \biggr] \cT_{q\bq}^{(0)} (\Phi_2) \,.
\end{align}
The $x_c$ dependence, which is a parameter of the subtraction formalism, must cancel between the regulated $gg$ virtual and $ggg$ real contributions and the $q\bq$ virtual and $gq\bq$ real contributions, and will provide a check on the calculation.

These matrix elements can be analytically integrated, which we do in the next section when we describe the total calculation of $C_3^{(2)}$.

%%%%%%%%%%%%%%%%%%%%%%%%%%%%%%%%%%%%%%%%%%%%%%%%%%%%%%%%%%%%%%%%%%%%%%%%%%%%%%%%%%%%%%%%%%%%%%%%%%%%%%%%%
\section{Calculation of $C_3^{(2)}$}
\label{sec:calc}
%%%%%%%%%%%%%%%%%%%%%%%%%%%%%%%%%%%%%%%%%%%%%%%%%%%%%%%%%%%%%%%%%%%%%%%%%%%%%%%%%%%%%%%%%%%%%%%%%%%%%%%%%

To calculate $C_3^{(2)}$, we must take the regulated real and virtual matrix elements determined in the previous sections and integrate them over the phase space against the measurement functions.

%%%%%%%%%%%%%%%%%%%%%%%%%%%%%%%%%%%%%%%%%%%%%%%%%%%%%%%%%%%%%%%%%%%%%%%%%%%%%%%%%%%%%%%%%%%%%%%%%%%%%%%%%
\subsection{The Regulated Virtual Contributions}
\label{subsec:regvirt}
%%%%%%%%%%%%%%%%%%%%%%%%%%%%%%%%%%%%%%%%%%%%%%%%%%%%%%%%%%%%%%%%%%%%%%%%%%%%%%%%%%%%%%%%%%%%%%%%%%%%%%%%%

The regulated virtual matrix elements in \eqs{VSgg}{VSqq} are integrated over the 2 particle phase space as shown in \eq{bornrealSn}, with $\Sym_{a_1 a_2} \cT^{(0)}_{a_1 a_2}$ replaced by $V^S_{a_1 a_2}$.  The evaluation of the virtual contributions follows closely the calculation of $C_2^{(1)}$, which is given in \app{C2}.  The only difference is the angular integral over $\Delta R_{12}$ which returns a double logarithm of $R$ and is given in \eq{lnRsqint}, and integrals over $x_1, x_2$.  There are three integrals needed for each channel:
\begin{align}
k^{(1)}_{a_1 a_2} &\equiv \int_0^\infty \df x_1 \df x_2 \, \frac{1}{(x_1 + x_2)^2} \, \Sym_{a_1 a_2} \langle \hat{P}_{a_1 a_2} (z) \rangle \bigl[ \theta(x_1 < 1) \theta(x_2 < 1) - \theta(x_1 + x_2 < 1) \bigr] \,, \nn \\
k^{(2)}_{a_1 a_2} &\equiv \int_0^\infty \df x_1 \df x_2 \, \frac{1}{(x_1 + x_2)^2} \, \Sym_{a_1 a_2} \langle \hat{P}_{a_1 a_2} (z) \rangle \, \ln (x_1 + x_2) \nn \\
& \qquad\qquad\qquad\qquad\qquad\qquad \times \bigl[ \theta(x_1 < 1) \theta(x_2 < 1) - \theta(x_1 + x_2 < 1) \bigr] \,, \nn \\
k^{(3)}_{a_1 a_2} &\equiv \int_0^\infty \df x_1 \df x_2 \, \frac{1}{(x_1 + x_2)^2} \, \Sym_{a_1 a_2} \langle \hat{P}_{a_1 a_2} (z) \rangle \, \ln \frac{x_1 x_2}{(x_1 + x_2)^2} \nn \\
& \qquad\qquad\qquad\qquad\qquad\qquad \times \bigl[ \theta(x_1 < 1) \theta(x_2 < 1) - \theta(x_1 + x_2 < 1) \bigr] \,.
\end{align}
The values of these integrals are, for the $gg$ channel:
\begin{align}
k^{(1)}_{gg} &= -\frac{1}{72} \, C_A (131 - 12\pi^2 - 132 \ln 2) \,, \nn \\
k^{(2)}_{gg} &= \frac{1}{432} \, C_A (-811 + 822 \ln 2 + 396 \ln^2 2 + 108 \zeta_3) \,, \nn \\
k^{(3)}_{gg} &= \frac{1}{216} \, C_A (1601 - 66\pi^2 - 822 \ln 2 - 396 \ln^2 2 - 540 \zeta_3) \,,
\end{align}
and for the $q\bq$ channel, adding in the flavor sum:
\begin{align}
k^{(1)}_{q\bq} &= -\frac{1}{36} \, T_F n_f (-23 + 24\ln 2) \,, \nn \\
k^{(2)}_{q\bq} &= \frac{1}{216} \, T_F n_f (163 - 174 \ln 2 - 72 \ln^2 2) \,, \nn \\
k^{(3)}_{q\bq} &= \frac{1}{108} \, T_F n_f (-305 + 12\pi^2 + 174 \ln 2 + 72 \ln^2 2) \,.
\end{align}
Note that $k^{(1)}$ is directly proportional to $C_2^{(1)}$ and $k^{(3)}$ is proportional to a clustering logarithm constant $s_2^{(1)}$ in Ref.~\cite{Stewart:2013faa}.  Using these results, the virtual contributions to the soft function are
\begin{align} \label{eq:C32virtterms}
\Delta S_{2,gg}^{(3)} &= \biggl(\frac{\as}{\pi}\biggr)^3 \ln \frac{\nu}{\pTcut} \ln^2 R^2 \, \Bigl\{ - \Bigl[ C_A^2 \ln x_c + \frac12 (\gamma_g + \beta_0 / 2) C_A \Bigr] k^{(1)}_{gg} + C_A^2 k^{(2)}_{gg} + C_A^2 k^{(3)}_{gg} \Bigr\} \,, \nn \\
\Delta S_{2,q\bq}^{(3)} &= \biggl(\frac{\as}{\pi}\biggr)^3 \ln \frac{\nu}{\pTcut} \ln^2 R^2 \, \Bigl\{ -\Bigl[ (2C_F - C_A) C_A \ln x_c + \frac12 (2\gamma_q - \gamma_g - \beta_0 / 2) C_A \Bigr] k^{(1)}_{q\bq} \nn \\
& \qquad\qquad\qquad\qquad\qquad\quad + (2C_F - C_A) C_A k^{(2)}_{q\bq} + C_F C_A k^{(3)}_{q\bq} \Bigr\} \,.
\end{align}
The contributions to $C_3^{(2)}$ can be extracted from these results by pulling out a factor of $(\as C_A/\pi)^3 \ln (\nu/\pTcut) \ln^2 R^2$.  We note that these contributions are the same for all $\kt$-type jet algorithms, as all algorithms in this group have the same phase space constraints with two particles in the final state.

%%%%%%%%%%%%%%%%%%%%%%%%%%%%%%%%%%%%%%%%%%%%%%%%%%%%%%%%%%%%%%%%%%%%%%%%%%%%%%%%%%%%%%%%%%%%%%%%%%%%%%%%%
\subsection{The Regulated Real Contributions}
\label{subsec:regreal}
%%%%%%%%%%%%%%%%%%%%%%%%%%%%%%%%%%%%%%%%%%%%%%%%%%%%%%%%%%%%%%%%%%%%%%%%%%%%%%%%%%%%%%%%%%%%%%%%%%%%%%%%%

%%%
\begin{figure}[ht!]
\begin{center}
\includegraphics[width=0.47\columnwidth]{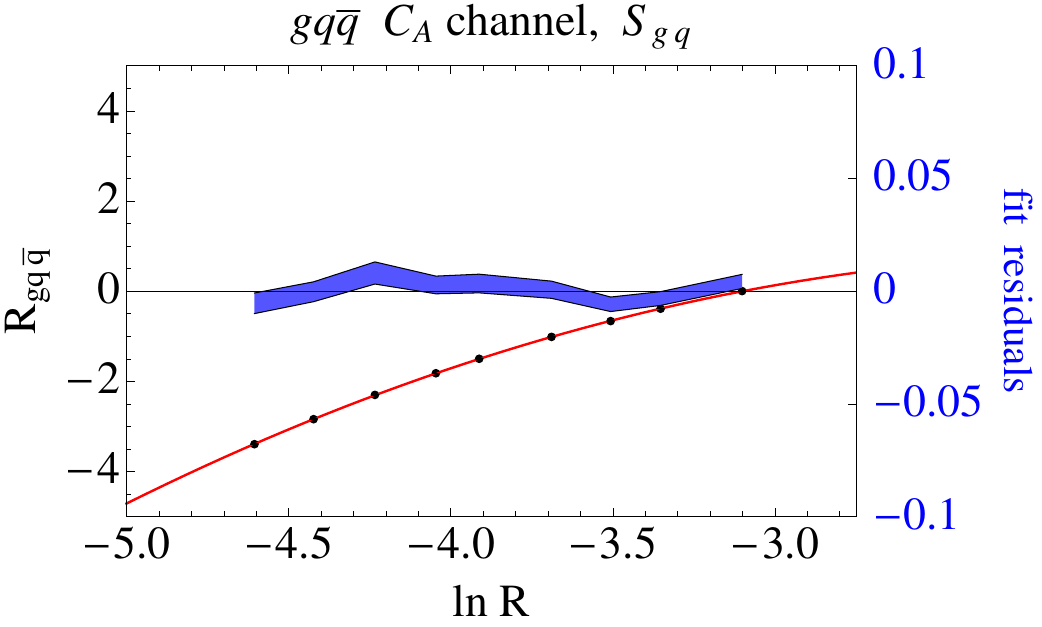} \hspace{3ex} \vspace{2ex}
\includegraphics[width=0.47\columnwidth]{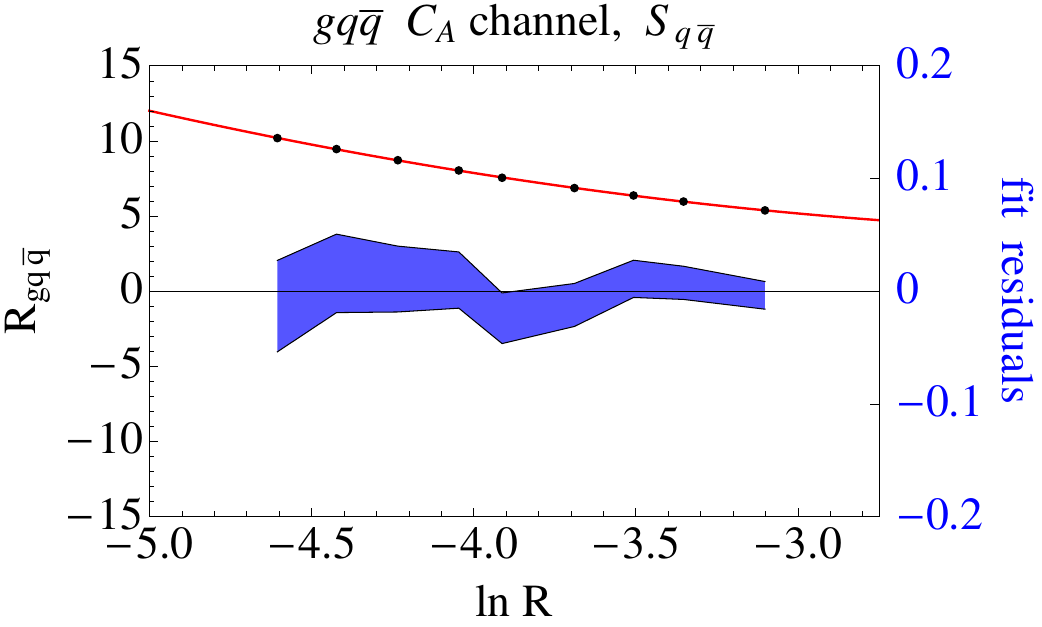}
\includegraphics[width=0.47\columnwidth]{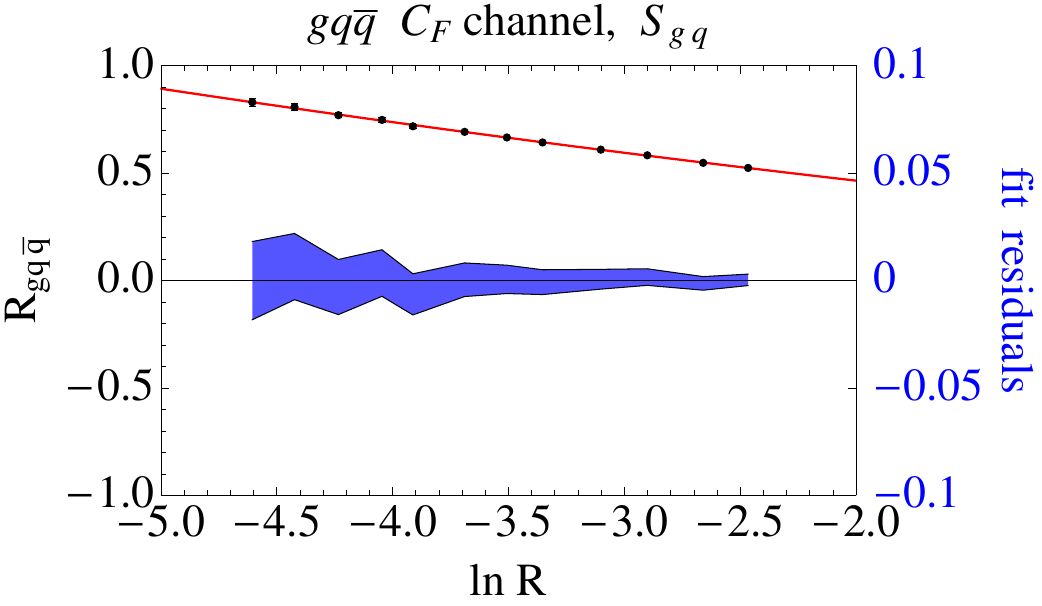} \hspace{3ex} \vspace{2ex}
\includegraphics[width=0.48\columnwidth]{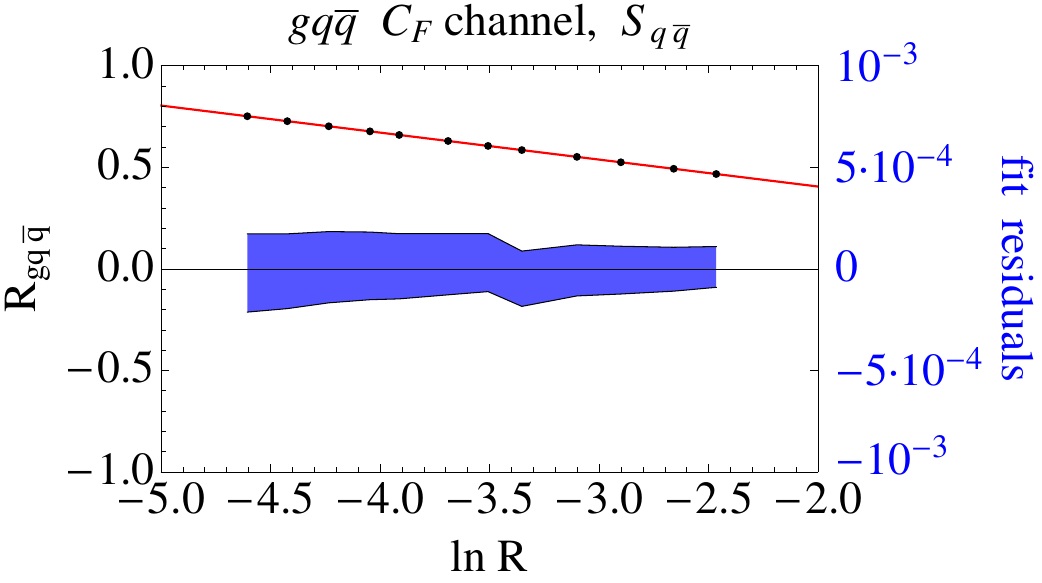}
\includegraphics[width=0.47\columnwidth]{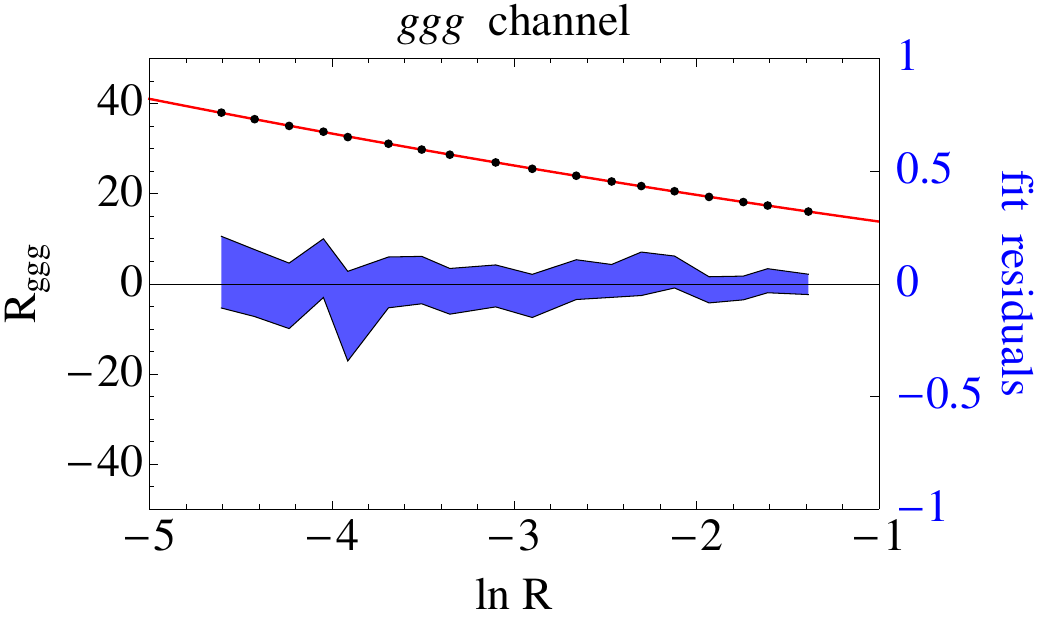}
\caption{Example fits to the regulated real emission contribution and fit residuals, for $R_c = 1.0$ and $x_c = 0.1$.  For each of the 5 distinct contributions, the black data points whose values and uncertainties come from the numerical calculation are plotted against the fit in red using the form in \eq{fitform}.  The residual difference is shown by the blue band, where the uncertainties on the data set the width of the band.  The blue numbers on the right of each plot set the vertical scale for the residuals.}
\label{fig:samplefits}
\end{center}
\end{figure}
%%%

The calculation of the regulated real contributions, $R_{ggg}$ and $R_{gq\bq}$, is performed numerically using the integration routine \texttt{VEGAS} in the \texttt{CUBA} library \cite{Hahn:2004fe}.  The calculation requires approximately $2\cdot 10^9$ events for a given set of parameters ($R, R_c, x_c$) sampled in the 7-dimensional phase space to achieve uncertainties on $C_3^{(2)}$ below $5\%$.

The calculation of $R_{ggg}$ and $R_{gq\bq}$ uses FKS-type subtractions.  For the $ggg$ channel symmetry implies there is only one unique sector, while for the $gq\bq$ channel there are two: $q\bq$ and $gq$ (which equals $g\bq$).  In the $gq\bq$ channel we also find it useful to divide the result into color factors $(C_A^2 n_f$ and $C_F C_A n_f)$, meaning there are four separate contributions to the $gq\bq$ result.  For each sector, the contribution to $C_3^{(2)}$ is determined through fits as a function of $\ln R$.  The fits are performed with $R$ values ranging between 0.01 and 0.25, using a fit to the quadratic function
\be \label{eq:fitform}
\textrm{fit form: } a_2 \ln^2 R + a_1 \ln R + a_0 \,.
\ee
For the $gq\bq$ channel we exclude larger $R$ values to reduce the impact of power corrections in $R$ on the fit.  The fits for $R_c = 1.0, x_c = 0.1$ across all the unique sectors is shown in \fig{samplefits}, and excellent fits are observed.

\begin{figure}[ht!]
\begin{center}
\includegraphics[width=0.6\columnwidth]{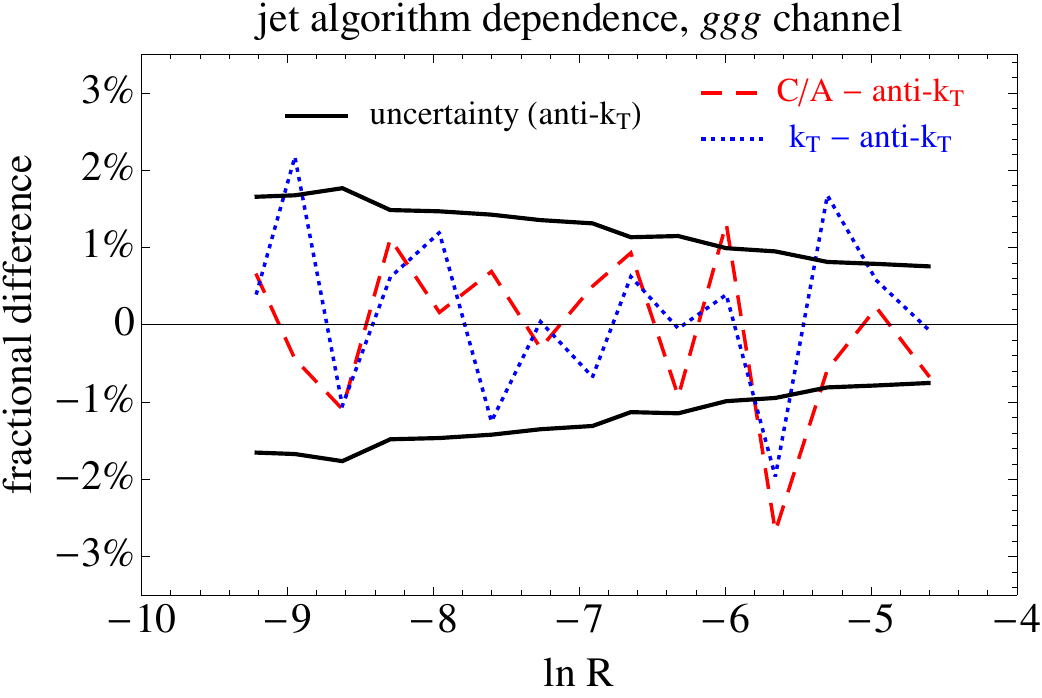}
\caption{Jet algorithm dependence in the calculation, shown for the $ggg$ channel as an example.  The fractional uncertainty for the anti-$\kt$ algorithm is shown in black, along with the percent difference in $R_{ggg}$ to the anti-$\kt$ result for the C/A (red, dashed) and $\kt$ (blue, dotted) algorithms.  The fact that the difference between algorithms lies within the uncertainty of the anti-$\kt$ result suggests that the value of $C_3^{(2)}$ is the same for the $\kt$-type algorithms.}
\label{fig:algdiff}
\end{center}
\end{figure}

We use the anti-$\kt$ algorithm as our primary jet algorithm, but in \fig{algdiff} we investigate the jet algorithm dependence of the calculation.  Since the virtual contributions are the same for all $\kt$-type algorithms, the algorithm dependence of $C_3^{(2)}$ is probed through the regulated real contributions.  Using $R_{ggg}$ as an example, we find that the differences between algorithms are within the uncertainties of the calculation.  This suggests that $C_3^{(2)}$ is the same for the $\kt$-type algorithms, and hence that the coefficient that we will extract is (at least somewhat) universal.
%%%
%%%

The coefficient $a_2$ in \eq{fitform} contributes to $C_3^{(2)}$, while the other coefficients are not used\footnote{Because the matrix elements are only valid in the triple collinear limit, the values of the subleading clustering logarithms may receive (arbitrarily) large corrections outside the collinear limit.  For example for $C_3^{(1)}$ the matrix elements needed are those with only one pair of collimated partons.}.  The entire fit procedure is performed over a grid of $R_c = \{0.2, 0.5, 1.0\}$ and $x_c = \{0.1, 0.3, 1.0\}$.  The variation in $R_c$ checks that the $C_3^{(2)}$ contribution is independent of its value, while the variation in $x_c$ must cancel with the integrated subtractions.  Indeed, we find this to be the case, and the final $C_3^{(2)}$ value (adding the real and virtual results together) for the 9 fits over different $R_c$ and $x_c$ values are very consistent with each other\footnote{The size of the canceling $x_c$ dependent terms in the regulated real and virtual contributions is large relative to the final value of $C_3^{(2)}$, so their cancellation is a robust check on the calculation.}.

%%%%%%%%%%%%%%%%%%%%%%%%%%%%%%%%%%%%%%%%%%%%%%%%%%%%%%%%%%%%%%%%%%%%%%%%%%%%%%%%%%%%%%%%%%%%%%%%%%%%%%%%%
\subsection{Results and Impact on the $H + 0$-jet Cross Section}
\label{subsec:results}
%%%%%%%%%%%%%%%%%%%%%%%%%%%%%%%%%%%%%%%%%%%%%%%%%%%%%%%%%%%%%%%%%%%%%%%%%%%%%%%%%%%%%%%%%%%%%%%%%%%%%%%%%

\begin{figure}[ht!]
\begin{center}
\includegraphics[width=0.47\columnwidth]{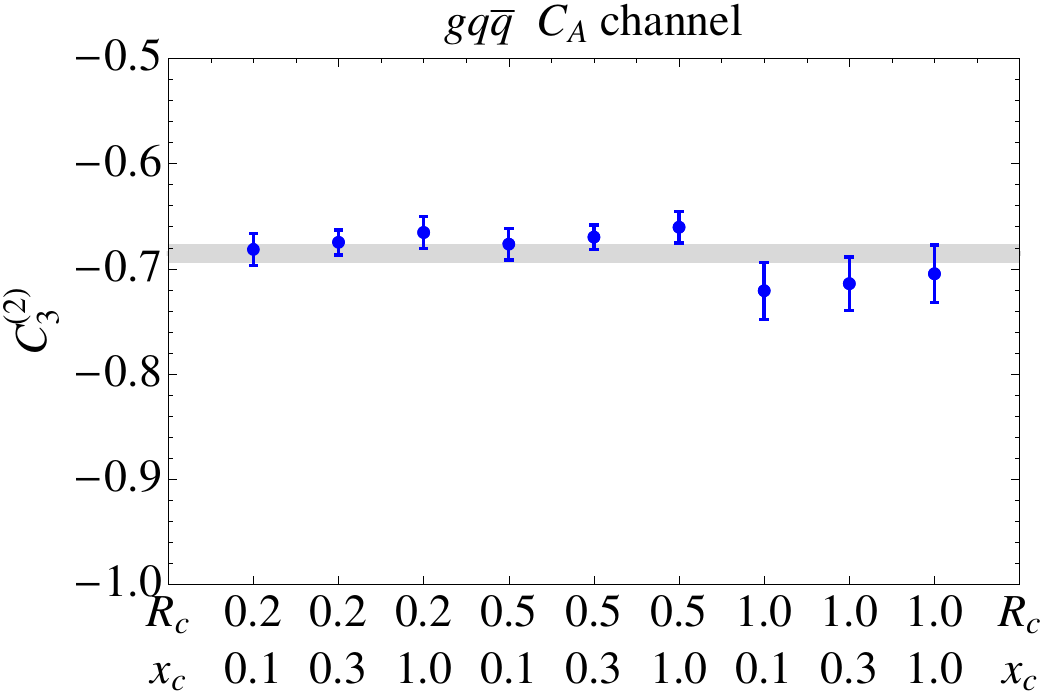} \hspace{3ex}  \vspace{2ex}
\includegraphics[width=0.47\columnwidth]{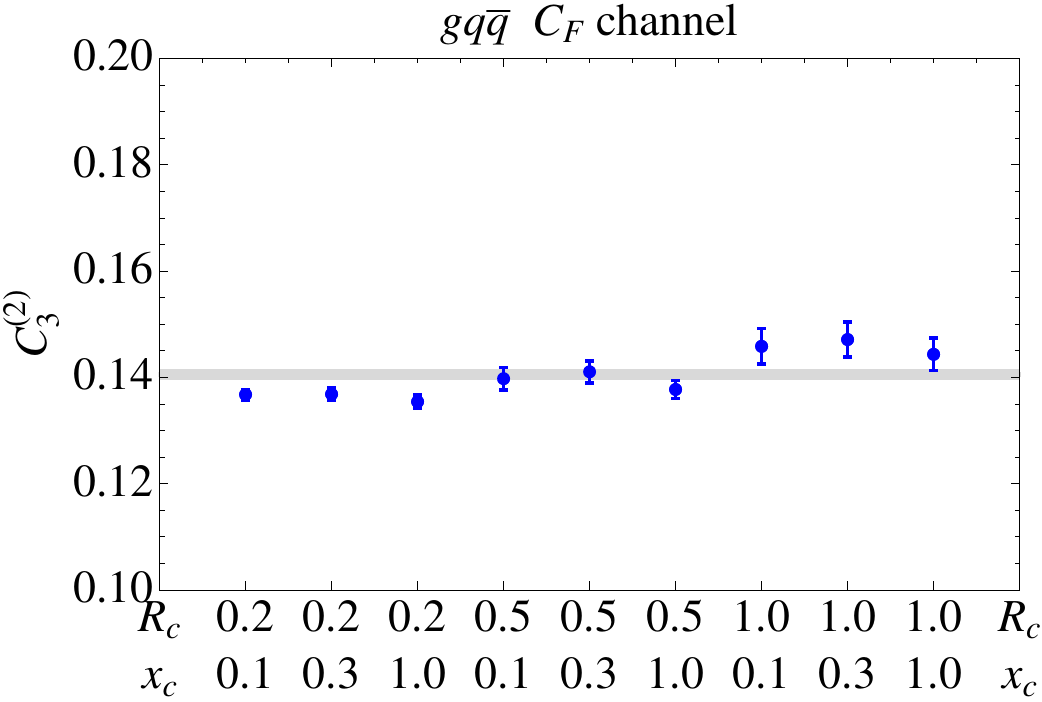}
\includegraphics[width=0.47\columnwidth]{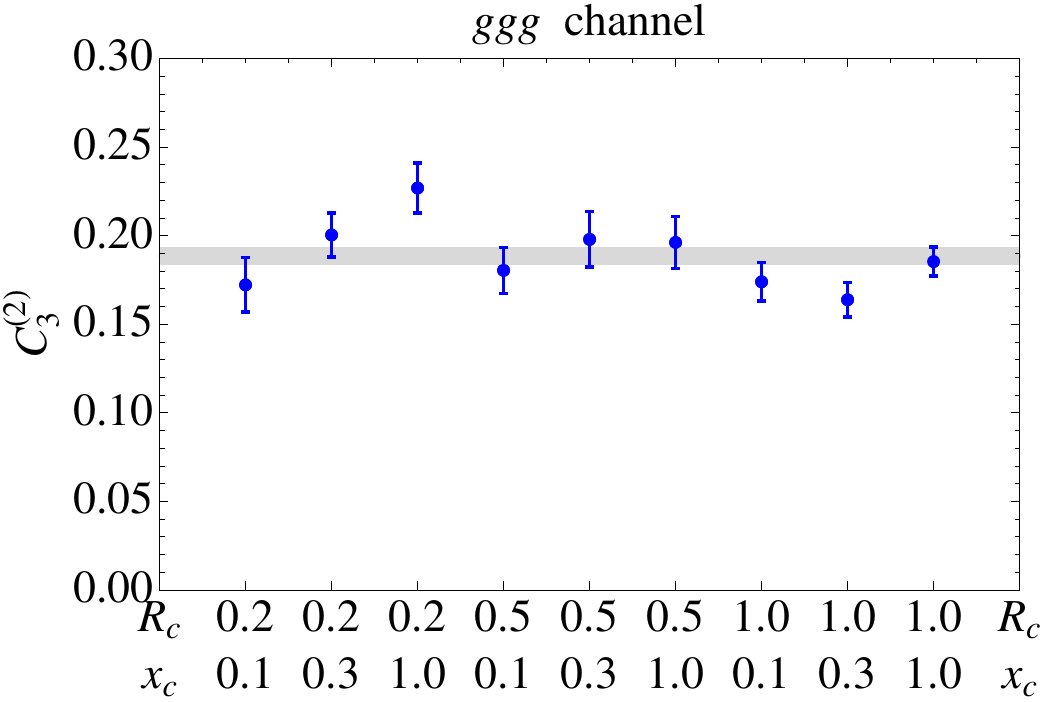}
\caption{Final results for the $C_3^{(2)}$ values in the 3 different channels.  The points are a combination of the fits to the regulated real emission, described in \subsec{regreal}, and the virtual contributions given in \eq{C32virtterms}.  The results are shown for various $R_c$ and $x_c$ values, and the coefficients are independent of these cut parameters.  The uncertainties are set entirely by the fits to the real emission terms, and the gray band shows the average value (with uncertainty).}
\label{fig:C3fits}
\end{center}
\end{figure}

With the results for the regulated real and virtual contributions in hand, we can combine them to determine $C_3^{(2)}$.  We add the analytically known virtual terms from \eq{C32virtterms} to the fits to the real terms and obtain 9 independent evaluations of $C_3^{(2)}$ (one for each $R_c$ and $x_c$ value).  The final result for $C_3^{(2)}$ is taken as the statistical average of these 9 determinations.  In \fig{C3fits}, we show the value of $C_3^{(2)}$ for the $ggg$ and $gq\bq$ channels for each fit as well as the averaged result.
%%%
%%%

We can evaluate the $ggg$ real + $gg$ virtual and $gq\bq$ + $q\bq$ virtual channels separately, and further divide the $gq\bq + q\bq$ results into $C_F C_A n_f$ and $C_A^2 n_f$ color channels.  The results from each channel, and the total contribution, are
\begin{align} \label{eq:C3final}
C_{3,ggg}^{(2)} &= 0.8889 \pm 0.0052 \,, \nn \\
C_{3,gq\bq, C_F}^{(2)} &= 0.1405\pm 0.0011 \,, \nn \\
C_{3,gq\bq, C_A}^{(2)} &= -0.5913 \pm 0.0091 \,, \nn \\
C_3^{(2)} &= C_{3,ggg}^{(2)} + C_{3,gq\bq, C_F}^{(2)} + C_{3,gq\bq, C_A}^{(2)} \nn \\
&= 0.438 \pm 0.011 \,.
\end{align}
Recall that $C_2^{(1)} \approx -2.49$, meaning the contribution of $C_3^{(2)}$ is much smaller.  Indeed, using the parameters $\{R = 0.4,\, \pTcut = 25 \GeV\}$ and $\{R = 0.5,\, \pTcut = 30 \GeV\}$ to determine the multiplicative factor to the resummed cross section, given in \eq{Undef}, we find
\begin{align}
U^{(2)}_{\rm clus} (0.4, 25 \GeV) - 1 &= 0.1520  \,, & U^{(3)}_{\rm clus} (0.4, 25 \GeV) - 1 &= (6.36 \pm 0.15) \cdot 10^{-3} \,, \nn \\
U^{(2)}_{\rm clus} (0.5, 30 \GeV) - 1 &= 0.0928  \,, & U^{(3)}_{\rm clus} (0.5, 30 \GeV) - 1 &= (2.91 \pm 0.07) \cdot 10^{-3} \,.
\end{align}
This means the leading $\ord{\as^3}$ clustering terms have an impact that is approximately 3\% of the leading $\ord{\as^2}$ clustering terms.  For two phenomenological points we give the cross section with and without $C_3^{(2)}$ (using $m_H = 125$ GeV and $E_{\rm cm} = 8$ TeV, and the NNLL$'$+NNLO result as our baseline):
\pagebreak
\begin{align}
\sigma_0 (\pTcut, R) & & \textrm{no } C_3^{(2)} & & \textrm{with } C_3^{(2)} \nn \\
R = 0.4 ,\, \pTcut = 25 \GeV: & & 12.67 \textrm{ [pb]} & & 12.75 \textrm{ [pb]} \nn \\
R = 0.5 ,\, \pTcut = 30 \GeV: & & 13.85 \textrm{ [pb]} & & 13.88 \textrm{ [pb]}
\end{align}

Although we have found that at $\ord{\as^3}$ the leading $\ln^2 R$ clustering logarithms are small, it would be interesting to determine the complete $\ord{\as^3}$ clustering corrections through a $H + 2$-jet NLO calculation.  Such a calculation is most powerful if it not only determines the numerical size of the correction for phenomenological ranges of parameters but also extracts the $\ord{\as^3}$ rapidity anomalous dimension contributions.  This would be an important part of extending the resummation for $H + 0$ jets to N$^3$LL, as the anomalous dimensions connected to the global veto contributions may be simpler to determine (due to the lack of a clustering algorithm).

$\\$
\break
\emph{Note Added:}
$\\$
The original version of this calculation neglected a contribution to $C_3^{(2)}$ from renormalization.  In that version, the virtual matrix elements in \eq{IC} were renormalized at the scale $\mu^2 = s_{12}$ instead of the scale $(\pTcut)^2$.  We have restored the UV pole (which is subsequently removed), and properly kept the resulting finite terms.  This is straightforwardly propagated through the calculation, and results in an additional contribution to $C_3^{(2)}$ equal to $(-\beta_0 / 8C_A) C_2^{(1)} = 0.7945$.  The current calculation reflects this term, including in \eq{C3final} and the following expressions, and our conclusions and the phenomenological statements are unaffected.  We thank the authors of Ref.~\cite{Dasgupta:2014yra} for discussions that clarified this issue.  The color decomposition of our main result, which may be more easily compared to the results of Ref.~\cite{Dasgupta:2014yra}, is $C_3^{(2)} C_A^3 = 0.889 C_A^3 + 0.379 C_F C_A T_F n_f - 0.611 C_A^2 T_F n_f - 0.118 C_A (T_F n_f)^2$.

%%%%%%%%%%%%%%%%%%%%%%%%%%%%%%%%%%%%%%%%%%%%%%%%%%%%%%%%%%%%%%%%%%%%%%%%%%%%%%%%%%%%%%%%%%%%%%%%%%%%%%%%%
\section{Conclusions}
\label{sec:conclusions}
%%%%%%%%%%%%%%%%%%%%%%%%%%%%%%%%%%%%%%%%%%%%%%%%%%%%%%%%%%%%%%%%%%%%%%%%%%%%%%%%%%%%%%%%%%%%%%%%%%%%%%%%%

In this work we have calculated the leading $\ord{\as^3}$ clustering logarithms for the jet vetoed cross section.  Clustering effects from the jet algorithm start at $\ord{\as^2}$, and those terms are numerically important.  We have calculated the most important terms in the $\ord{\as^3}$ correction, those with the form $\as^3 \ln^2 R \ln m_H / \pTcut$, and find that they are numerically less important than the $\ord{\as^2}$ terms.  This brings the higher order clustering effects under better control, and is the first step in determining the complete clustering effects that take part in the N$^3$LL resummation of the $H + 0$-jet cross section.

In addition to the improved description of jet vetoed cross sections, this work combines several techniques to perform the calculation.  What is a naively N$^3$LO calculation is reduced to an NLO calculation through the combined factorization properties of QCD and SCET.  Subtraction techniques for NLO calculations are then used to perform the calculation.  Our success suggests that these techniques may be useful in other SCET calculations, and from a pedagogical perspective it would be interesting to explore how SCET interfaces with fixed-order subtractions beyond NLO.  In \sec{subtractions} we have seen an example of how soft, collinear, and collinear-soft subtractions at NLO are given by SCET matrix elements.  This is expected, since subtractions and SCET matrix elements are based on the same singular limit of QCD amplitudes.  At NNLO, the factorization properties and organization methods of SCET may help organize the larger set of subtraction terms.

%%%%%%%%%%%%%%%%%%%%%%%%%%%%%%%%%%%%%%%%%%%%%%%%%%%%%%%%%%%%%%%%%%%%%%%%%%%%%%%%%%%%%%%%%%%%%%%%%%%%%%%%%
\section*{Acknowledgments}
\label{sec:acks}
%%%%%%%%%%%%%%%%%%%%%%%%%%%%%%%%%%%%%%%%%%%%%%%%%%%%%%%%%%%%%%%%%%%%%%%%%%%%%%%%%%%%%%%%%%%%%%%%%%%%%%%%%

We thank Christian Bauer, Frank Petriello, Maximilian Stahlhofen, Iain Stewart, Frank Tackmann, and Saba Zuberi for helpful discussions and comments on the manuscript.  This work was supported in part by the Department of Energy Early Career Award with Funding Opportunity No. DE-PS02-09ER09-26, the Office of High Energy Physics of the U.S. Department of Energy under the Contract DE-AC02-05CH11231, and the US National Science Foundation, grant NSF-PHY-0705682, the LHC Theory Initiative.  This research used resources of the National Energy Research Scientific Computing Center, which is supported by the Office of Science of the U.S. Department of Energy under Contract No. DE-AC02-05CH11231.

\appendix

%%%%%%%%%%%%%%%%%%%%%%%%%%%%%%%%%%%%%%%%%%%%%%%%%%%%%%%%%%%%%%%%%%%%%%%%%%%%%%%%%%%%%%%%%%%%%%%%%%%%%%%%%
\section{$1\to2$ and $1\to3$ Splitting Functions}
\label{app:splittingfunctions}
%%%%%%%%%%%%%%%%%%%%%%%%%%%%%%%%%%%%%%%%%%%%%%%%%%%%%%%%%%%%%%%%%%%%%%%%%%%%%%%%%%%%%%%%%%%%%%%%%%%%%%%%%

The $1\to2$ gluon-initiated splitting functions and their polarization averages are
\begin{align} \label{eq:1to2splitting}
g\to gg &: \nn \\
\hat{P}_{gg}^{\mu \nu} &= 2C_A \biggl[ -g^{\mu\nu} \biggl(\frac{z}{1-z} + \frac{1-z}{z} \biggr) - 2(1-\e) z(1-z) \frac{k_{\perp1}^\mu k_{\perp1}^\nu}{k_{\perp1}^2} \biggr] \,, \nn \\
\langle \hat{P}_{gg} \rangle &= 2C_A \biggl[ \frac{z}{1-z} + \frac{1-z}{z} + z(1-z) \biggr] \,, \nn \\
g\to q\bq &: \nn \\
\hat{P}_{q\bq}^{\mu \nu} &= T_F \biggl[ -g^{\mu\nu} + 4z(1-z) \frac{k_{\perp1}^\mu k_{\perp1}^\nu}{k_{\perp1}^2} \biggr] \,, \nn \\
\langle \hat{P}_{q\bq} \rangle &= T_F \biggl[ 1 - \frac{2z(1-z)}{1-\e} \biggr] \,.
\end{align}

The $1\to3$ gluon-initiated splitting functions and their polarization averages are
\begin{align} \label{eq:1to3splitting}
g\to ggg &: \nn \\
\hat{P}_{ggg}^{\mu\nu} &= C_A^2 \biggl\{ \frac{(1-\e)}{4s_{12}^2} \biggl[ -g^{\mu\nu} t_{12,3}^2 + 16 s_{123} \frac{z_1^2 z_2^2}{z_3 (1-z_3)} \biggl( \frac{k_{\perp 2}}{z_2} - \frac{k_{\perp 1}}{z_1} \biggr)^\mu \biggl( \frac{k_{\perp 2}}{z_2} - \frac{k_{\perp 1}}{z_1} \biggr)^\nu \, \biggr] \nn \\
& \qquad\qquad - \frac34 (1-\e) g^{\mu\nu} + \frac{s_{123}}{s_{12}} g^{\mu\nu} \frac{1}{z_3} \biggl[ \frac{2(1-z_3) + 4z_3^2}{1-z_3} - \frac{1 - 2z_3(1-z_3)}{z_1(1 - z_1)} \biggr] \nn \\
& \qquad\qquad + \frac{s_{123}(1-\e)}{s_{12} s_{13}} \biggl[ 2z_1 \biggl( k_{\perp 2}^\mu k_{\perp 2}^\nu \frac{1-2z_3}{z_3(1-z_3)} + k_{\perp 3}^\mu k_{\perp 3}^\nu \frac{1 - 2z_2}{z_2(1-z_2)} \biggr) \nn \\
& \qquad\qquad\qquad + \frac{s_{123}}{2(1-\e)} g^{\mu\nu} \biggl( \frac{4z_2 z_3 + 2z_1(1 - z_1) - 1}{(1-z_2)(1-z_3)} - \frac{1 - 2z_1(1-z_1)}{z_2 z_3} \biggr) \nn \\
& \qquad\qquad\qquad + \bigl( k_{\perp 2}^\mu k_{\perp 3}^\nu + k_{\perp 3}^\mu k_{\perp 2}^\nu \bigr) \biggl( \frac{2z_2(1-z_2)}{z_3(1-z_3)} - 3 \biggr) \biggr] \biggr\} + (5 \textrm{ permutations}) \,, \nn \\
\langle \hat{P}_{ggg} \rangle &= C_A^2 \biggl\{ \frac{(1-\e)}{4s_{12}^2} t_{12,3}^2 + \frac34 (1-\e) + \frac{s_{123}}{s_{12}} \biggl[ 4\frac{z_1 z_2 - 1}{1 - z_3} + \frac{z_1 z_2 - 2}{z_3} + \frac32 + \frac52 z_3 \nn \\
& \qquad + \frac{(1 - z_3(1 - z_3))^2}{z_3 z_1 (1 - z_1)} \biggr] + \frac{s_{123}^2}{s_{12}s_{13}} \biggl[ \frac{z_1 z_2 (1-z_1)(1 - 2z_3)}{z_3(1-z_3)} + z_2 z_3 - 2 + \frac{z_1 (1+2z_1)}{2} \nn \\
& \qquad + \frac{1 + 2z_1(1 + z_1)}{2(1-z_2)(1-z_3)} + \frac{1 - 2z_1(1-z_1)}{2z_2z_3} \biggr] \biggr\} + (5 \textrm{ permutations}) \,.
\end{align}
\begin{align}
g\to gq\bq &: \nn \\
\hat{P}_{gq\bq}^{\mu\nu} &= C_F T_F \biggl\{ -\frac12 g^{\mu\nu} \biggl[ -2 + \frac{2s_{123}s_{23} + (1-\e)(s_{123} - s_{23})^2}{s_{12}s_{13}} \biggr] \nn \\
& \qquad\qquad + \frac{2s_{123}}{s_{12}s_{13}} \bigl( k_{\perp 2}^\mu k_{\perp 3}^\nu + k_{\perp 3}^\mu k_{\perp 2}^\nu - (1-\e) k_{\perp 1}^\mu k_{\perp 1}^\nu \bigr) \biggr\} \nn \\
& + C_A T_F \biggl\{ \frac{s_{123}}{4s_{23}^2} \biggl[ g^{\mu\nu} \frac{t_{23,1}^2}{s_{123}} - 16 \frac{z_2^2 z_3^2}{z_1(1-z_1)} \biggl( \frac{k_{\perp 2}}{z_2} - \frac{k_{\perp 3}}{z_3} \biggr)^\mu \biggl( \frac{k_{\perp 2}}{z_2} - \frac{k_{\perp 3}}{z_3} \biggr)^\nu \, \biggr] \nn \\
& \qquad\qquad + \frac{s_{123}}{4s_{12}s_{13}} \biggl[ 2s_{123} g^{\mu\nu} - 4\bigl( k_{\perp 2}^\mu k_{\perp 3}^\nu + k_{\perp 3}^\mu k_{\perp 2}^\nu - (1-\e) k_{\perp 1}^\mu k_{\perp 1}^\nu \bigr) \biggr] \nn \\
& \qquad\qquad - \frac14 g^{\mu\nu} \biggl[ -(1-2\e) + 2\frac{s_{123}}{s_{12}} \frac{1 - z_3}{z_1(1-z_1)} + 2\frac{s_{123}}{s_{23}} \frac{1 - z_1 + 2z_1^2}{z_1(1-z_1)} \biggr] \nn \\
& \qquad\qquad + \frac{s_{123}}{4s_{12}s_{23}} \biggl[ -2s_{123}g^{\mu\nu} \frac{z_2(1-2z_1)}{z_1(1-z_1)} - 16k_{\perp 3}^\mu k_{\perp 3}^\nu \frac{z_2^2}{z_1(1-z_1)} + 8(1-\e) k_{\perp 2}^\mu k_{\perp 2}^\nu \nn \\
& \qquad\qquad \qquad + 4(k_{\perp 2}^\mu k_{\perp 3}^\nu + k_{\perp 3}^\mu k_{\perp 2}^\nu) \biggl( \frac{2 z_2(z_3 - z_1)}{z_1(1-z_1)} + (1-\e) \biggr) \biggr] \biggr\} + (2\leftrightarrow3) \,, \nn \\
\langle \hat{P}_{gq\bq} \rangle &= \frac12 C_F T_F \biggl\{ -2 - (1-\e)\biggl(\frac{s_{23}}{s_{12}} + \frac{s_{23}}{s_{13}} \biggr) + 2 \frac{s_{123}^2}{s_{12}s_{13}} \biggl(1 + z_1^2 - \frac{z_1 + 2z_2 z_3}{1-\e} \biggr) \nn \\
& \qquad - \frac{s_{123}}{s_{12}} \biggl( 1 + 2z_1 + \e - \frac{2}{1-\e}(z_1 + z_2) \biggr) - \frac{s_{123}}{s_{13}} \biggl( 1 + 2z_1 + \e - \frac{2}{1-\e}(z_1 + z_3) \biggr) \biggr\} \nn \\
& + C_A T_F \biggl\{ - \frac{t_{23,1}^2}{4s_{23}^2} + \frac{s_{123}^2}{2s_{13}s_{23}} z_3 \biggl[ \frac{(1-z_1)^3 - z_1^3}{z_1(1-z_1)} - \frac{2z_3(1 - z_3 - 2z_1 z_2)}{(1-\e) z_1(1 - z_1)} \biggr] \nn \\
& \qquad + \frac{s_{123}}{2s_{13}} (1-z_2) \biggl[ 1 + \frac{1}{z_1(1-z_1)} - \frac{2z_2(1-z_2)}{(1-\e)z_1(1-z_1)} \biggr] \nn \\
& \qquad + \frac{s_{123}}{2s_{23}} \biggl[ \frac{1 + z_1^3}{z_1(1-z_1)} + \frac{z_1(z_3 - z_2)^2 - 2z_2 z_3 (1 + z_1)}{(1-\e)z_1(1-z_1)} \biggr] \nn \\
& \qquad - \frac14 + \frac{\e}{2} - \frac{s_{123}^2}{2s_{12}s_{13}} \biggl(1 + z_1^2 - \frac{z_1 + 2z_2 z_3}{1-\e} \biggr) \biggr\} + (2 \leftrightarrow 3) \,.
\end{align}
In the $g\to gq\bq$ splitting functions, $g = 1$, $q = 2$, $\bq = 3$.  Additionally
\be
t_{ij,k} \equiv 2\frac{z_i s_{jk} - z_j s_{ik}}{z_i + z_j} + \frac{z_i - z_j}{z_i + z_j} s_{ij} \,.
\ee

%%%%%%%%%%%%%%%%%%%%%%%%%%%%%%%%%%%%%%%%%%%%%%%%%%%%%%%%%%%%%%%%%%%%%%%%%%%%%%%%%%%%%%%%%%%%%%%%%%%%%%%%%
\section{Integrating the Subtractions}
\label{app:integratedsubtractions}
%%%%%%%%%%%%%%%%%%%%%%%%%%%%%%%%%%%%%%%%%%%%%%%%%%%%%%%%%%%%%%%%%%%%%%%%%%%%%%%%%%%%%%%%%%%%%%%%%%%%%%%%%

In this appendix we detail the integration of the subtraction terms defined in \sec{subtractions}.  As these subtractions are closely related to the usual FKS ones, the integrated subtractions will have a very similar form.  The primary difference comes from the fact that the matrix elements and subtractions are defined in the small-angle limit of the final state.

The integrated subtraction is derived in the following way.  For each subtraction, we take the relevant limit of the real matrix element and factorize the phase space integration into the Born phase space multiplied by the real emission phase space.  The integral of the subtraction factor over this real emission phase space in $d$ dimensions gives the integrated subtraction.

For the soft and collinear-soft subtractions the phase space factorization follows the same steps, so we show them for the soft subtraction.  The radiative phase space is defined by $x_3$, $y_{13}$, and $\phi_{13}$, while the Born phase space is given by $x_1$, $x_2$, $y_{12}$, and $\phi_{12}$.  We want to factor off the Born contribution explicitly, absorbing the remainder into the integrated subtractions.  Recalling \eq{bornrealSn} and keeping only the $\ln \nu/\pTcut$ term from the rapidity divergence, we can split off the Born phase space and obtain
\begin{align} \label{eq:Tggg3softlimit}
&\int \df \Phi_3 \, \cT^{(0)}_{ggg} (\Phi_3) \Delta \cM^{(3)} (\Phi_3) \; \xrightarrow[3 \,\to\, \textrm{soft}]{} \; \biggl[ \int \df \Phi_2 \, \cT^{(0)}_{gg} (\Phi_2) \Delta \cM^{(2)} (\Phi_2) \biggr] \nn \\
& \qquad \times \frac{\as}{2\pi} \frac{e^{\gamma_E \e}}{\Gamma(1-\e)} \biggl(\frac{\mu}{\pTcut}\biggr)^{2\e} \int_0^\infty \df x_3 \, x_3^{1-2\e} \int_{-\infty}^\infty \df y_{k3} \int_{-\pi}^\pi \frac{\df\phi_{k3}}{2\pi} c_{\phi k3} \sum_{k,l} \wh{S}^3_{kl} \,,
\end{align}
where
\be
S^{i}_{kl} = (4\pi\as\mu^{2\e}) \frac{1}{(\pTcut)^2} \wh{S}^{i}_{kl} \,.
\ee
We rescale the other subtractions terms similarly and denote them with hat, $\wh{C}$ and $\wh{\CS}$.  \eq{Tggg3softlimit} implies that the integrated soft subtraction is, once we add in the phase space cut on the auxiliary $x_c$,
\be
I_S^{i,kl} = \frac{\as}{2\pi} \frac{e^{\gamma_E \e}}{\Gamma(1-\e)} \biggl(\frac{\mu}{\pTcut}\biggr)^{2\e} \int_0^\infty \df x_i \, x_i^{1-2\e} \int_{-\infty}^\infty \df y_{ik} \int_{-\pi}^\pi \frac{\df\phi_{ik}}{2\pi} c_{\phi ik} \, \wh{S}_{kl}^i \, \theta( x_i < x_c) \,.
\ee
For the case of the $1r$ or $2r$ soft gluon exchange, where the form matches the collinear-soft subtractions, we also add a cut on $R_c$.  Similarly, the integrated collinear-soft subtraction is
\be
I_{\CS}^{ij} = \frac{\as}{2\pi} \frac{e^{\gamma_E \e}}{\Gamma(1-\e)} \biggl(\frac{\mu}{\pTcut}\biggr)^{2\e} \int_0^\infty \df x_i \, x_i^{1-2\e} \int_{-\infty}^\infty \df y_{ij} \int_{-\pi}^\pi \frac{\df\phi_{ij}}{2\pi} c_{\phi ij} \, \wh{\CS}_{ij} \, \theta( x_i < x_c) \theta( \Delta R_{ij} < R_c) \,.
\ee
For the collinear subtraction, the Born matrix element is expressed in terms of momenta $k_1 + k_3$ (in the collinear limit) and $k_2$.  This means the $p_T$ fraction variables must be written in terms of $x_1 + x_3$ and $x_2$, while the angular variables are suitable as-is (the collinear splitting is parameterized by $y_{13}$ and $\phi_{13}$, the Born by $y_{12}$ and $\phi_{12}$).  We perform the change of variables
\begin{align}
&x_{13} \equiv x_1 + x_3 \,, \qquad z \equiv \frac{x_1}{x_1 + x_3} \,, \nn \\
&x_{13}\, \df x_{13} \df z = \df x_1 \df x_3 \,.
\end{align}
The variable $z$ parameterizes the collinear splitting.  In terms of these variables, the collinear limit of the real matrix element is
\begin{align}
&\int \df \Phi_3 \, \cT^{(0)}_{ggg} (\Phi_3) \Delta \cM^{(3)} (\Phi_3) \; \xrightarrow[1,3 \,\to\, \textrm{coll}]{} \; \biggl[ \int \df \Phi_2 \, \cT^{(0)}_{gg} (\Phi_2) \Delta \cM^{(2)} (\Phi_2) \biggr] \nn \\
& \qquad \times \frac{\as}{2\pi} \frac{e^{\gamma_E \e}}{\Gamma(1-\e)} \biggl(\frac{\mu}{\pTcut}\biggr)^{2\e} \int_0^1 \df z \, [z(1-z)]^{1-2\e} \int_{-\infty}^\infty \df y_{ij} \int_{-\pi}^\pi \frac{\df\phi_{ij}}{2\pi} c_{\phi ij} \, z_{13}^2 \, \wh{C}_{ij} \,,
\end{align}
Adding the cut on $R_c$, the integrated collinear subtraction is
\be \label{eq:ICform}
I_C^{ij} = \frac{\as}{2\pi} \frac{e^{\gamma_E \e}}{\Gamma(1-\e)} \biggl(\frac{\mu}{\pTcut x_{ij}}\biggr)^{2\e} \int_0^1 \df z \, [z(1-z)]^{1-2\e} \int_{-\infty}^\infty \df y_{ij} \int_{-\pi}^\pi \frac{\df\phi_{ij}}{2\pi} c_{\phi ij} \, z_{13}^2 \, \wh{C}_{ij} \, \theta( \Delta R_{ij} < R_c) \,.
\ee
Note that the Born matrix element is expressed in terms of the momenta $k_1 + k_3$ and $k_2$.  However, since we have written the phase space integration in terms of these momenta, they are really just dummy variables at this point, and this is why the collinear subtraction factors out cleanly.

%%%%%%%%%%%%%%%%%%%%%%%%%%%%%%%%%%%%%%%%%%%%%%%%%%%%%%%%%%%%%%%%%%%%%%%%%%%%%%%%%%%%%%%%%%%%%%%%%%%%%%%%%
\subsection{Soft Subtractions}
\label{app:integratedsoft}
%%%%%%%%%%%%%%%%%%%%%%%%%%%%%%%%%%%%%%%%%%%%%%%%%%%%%%%%%%%%%%%%%%%%%%%%%%%%%%%%%%%%%%%%%%%%%%%%%%%%%%%%%

There are two types of soft integrated subtractions: those which exchange the soft gluon between two collimated partons (e.g., 1 and 2) and those which exchange the soft gluon between one collimated parton and the rest of the event in a color coherent way (e.g., 1 and $r$).  In the first case, the integrated subtraction is
\begin{align}
I_S^{3,12} = \frac{\as}{2\pi} \frac{e^{\gamma_E \e}}{\Gamma(1-\e)} \biggl(\frac{\mu}{\pTcut}\biggr)^{2\e} (-\T_1 \cdot \T_2) \int_0^{x_c} \df x_3 \, x_3^{-1-2\e} \int_{-\infty}^\infty \df y_{13} \int_{-\pi}^\pi \frac{\df\phi_{13}}{2\pi} c_{\phi 13} \, \frac{2\Delta R_{12}^2}{\Delta R_{13}^2 \Delta R_{23}^2} \,.
\end{align}
The $p_T$ fraction $x_3$ can be integrated over directly, and the angular integrals are straightforward by using
\be
\Delta R_{23}^2 = (y_{13} - y_{12})^2 + (\phi_{13} - \phi_{12})^2 \,,
\ee
and integrating over $y_{13}$ first (which is independent of $\e$).  The result is
\be \label{eq:IS12}
I_S^{3,12} = \frac{\as}{2\pi} \frac{e^{\gamma_E\e}}{\Gamma(1-\e)} (-\T_1\cdot \T_2) \biggl( \frac{\mu}{\pTcut x_c \Delta R_{12}} \biggr)^{2\e} \biggl[ \frac{1}{\e^2} + \frac{\pi^2}{12} \biggr] \,.
\ee
This result is basically identical to the integrated FKS soft subtraction taken in the limit $\Delta R_{12} \ll 1$.  A similar soft function was calculated in Ref.~\cite{Ellis:2010rwa}.

\begin{figure}[!h]
\begin{center}
\includegraphics[width=0.5\columnwidth]{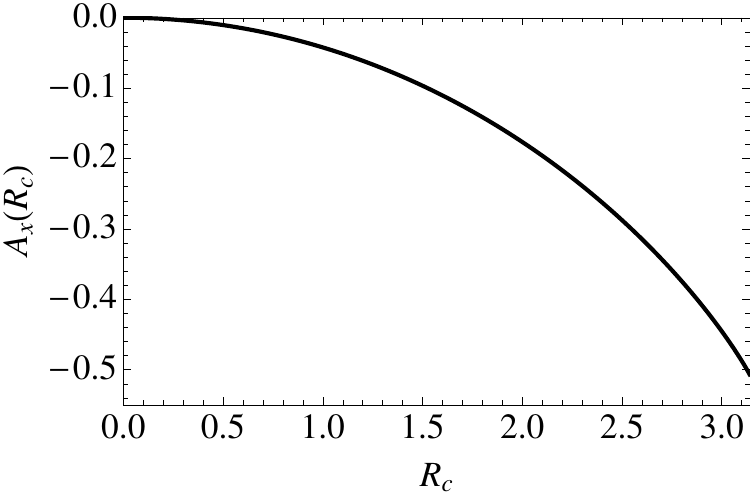}
\caption{The integral $A_x (R_c)$ defined in \eq{AxRc}.}
\label{fig:AxRc}
\end{center}
\end{figure}

The second type of soft subtraction matches the kinematic dependence of the collinear-soft subtraction.  For the collinear-soft subtraction we will impose cuts on $x_c$ and $R_c$, while for the soft subtraction we will only impose a cut on $x_c$.  The integrated soft subtraction is
\begin{align}
I_S^{3,1r} &= \frac{\as}{2\pi} \frac{e^{\gamma_E \e}}{\Gamma(1-\e)} \biggl(\frac{\mu}{\pTcut}\biggr)^{2\e} (-\T_1 \cdot \T_r) \int_0^{x_c} \df x_3 \, x_3^{-1-2\e} \nn \\
& \qquad \times \int_{-\infty}^\infty \df y_{13} \int_{-\pi}^\pi \frac{\df\phi_{13}}{2\pi} c_{\phi 13} \frac{2}{\Delta R_{13}^2} \,.
\end{align}
The collinear-soft subtraction has an additional constraint $\theta(\Delta R_{13} < R_c)$.  The angular integrals are again straightforward, but there is a constant term depending on $R_c$ whose functional form is difficult to obtain.  These integrals to $\ord{\e}$ are (with and without the $R_c$ constraint):
\begin{align}
\int_{-\infty}^\infty \df y_{13} \int_{-\pi}^\pi \frac{\df\phi_{13}}{2\pi} c_{\phi 13} \frac{2}{\Delta R_{13}^2} &= (2\pi)^{-2\e} \biggl[ -\frac{1}{\e} + \frac{\pi^2}{6} \e + 2\e A_\phi \biggr] \,, \nn \\
\int_{-\infty}^\infty \df y_{13} \int_{-\pi}^\pi \frac{\df\phi_{13}}{2\pi} c_{\phi 13} \frac{2}{\Delta R_{13}^2} \theta(\Delta R_{13} < R_c) &= R_c^{-2\e} \biggl[ -\frac{1}{\e} - 4\e A_x (R_c) \biggr] \,,
\end{align}
where
\be
A_\phi = -2\int_0^1 \frac{\df x}{x} \ln \biggl( \frac{\sin \pi x}{\pi x} \biggr) = \sum_{k=1}^{\infty} \Li_2 (1/k^2) = 2.31267895042751632\ldots \,,
\ee
and
\begin{align} \label{eq:AxRc}
A_x (R_c) &= \int_0^1 \frac{\df x}{x} \ln \biggl( \frac{\sin R_c x}{R_c x} \biggr) \frac{2}{\pi} \cos^{-1} x \nn \\
&= -\frac14 \sum_{k=1}^{\infty} \frac{R_c^2}{k^2 \pi^2} \, {}_4 F_3 \Bigl( \{1,1,1,3/2\} , \{2,2,2\}, \frac{R_c^2}{k^2 \pi^2} \Bigr) \,,
\end{align}
and is plotted in \fig{AxRc}.
%%%
%%%
This soft subtraction is therefore
\be \label{eq:IS1r}
I_S^{3,1r} = \frac{\as}{2\pi} \frac{e^{\gamma_E \e}}{\Gamma(1-\e)} \biggl(\frac{\mu}{\pTcut 2\pi x_c}\biggr)^{2\e} (-\T_1 \cdot \T_r) \biggl[ \frac{1}{2\e^2} - \frac{\pi^2}{12} - A_\phi \biggr] \,.
\ee

%%%%%%%%%%%%%%%%%%%%%%%%%%%%%%%%%%%%%%%%%%%%%%%%%%%%%%%%%%%%%%%%%%%%%%%%%%%%%%%%%%%%%%%%%%%%%%%%%%%%%%%%%
\subsection{Collinear Subtractions}
\label{app:integratedcoll}
%%%%%%%%%%%%%%%%%%%%%%%%%%%%%%%%%%%%%%%%%%%%%%%%%%%%%%%%%%%%%%%%%%%%%%%%%%%%%%%%%%%%%%%%%%%%%%%%%%%%%%%%%

The integrated collinear subtraction can be written from \eqs{C13}{ICform} as
\begin{align} \label{eq:IC13}
I_C^{13} &= \frac{\as}{2\pi} \frac{e^{\gamma_E \e}}{\Gamma(1-\e)} \biggl(\frac{\mu}{\pTcut x_{13}}\biggr)^{2\e} \int_0^1 \df z \, [z(1-z)]^{-2\e} \langle \hat{P}_{gg} (z) \rangle \nn \\
& \qquad \times \int_{-\infty}^\infty \df y_{13} \int_{-\pi}^\pi \frac{\df\phi_{13}}{2\pi} c_{\phi 13} \, \frac{2}{\Delta R_{13}^2} \, \theta( \Delta R_{13} < R_c) \,.
\end{align}
The angular integrals have already been carried out in the soft subtractions above, and the integral over $z$ is straightforward.  The result is
\be \label{eq:IC13gg}
I_C^{13} = \frac{\as}{2\pi} \frac{e^{\gamma_E \e}}{\Gamma(1-\e)} \biggl(\frac{\mu}{\pTcut x_{13} R_c}\biggr)^{2\e} \biggl[ \frac{2}{\e^2} C_A + \frac{2}{\e} \gamma_g^{(g)} + 2\gamma_g^{' \, (g)} + 8 A_x (R_c) C_A \biggr] \,.
\ee
In this case we have not included a symmetry factor for the $g\to gg$ splitting.  Otherwise, the result is very similar to the FKS integrated collinear subtraction, save for the different constant.  For $q\to qg$ splittings, the result is
\be \label{eq:IC13qg}
I_C^{qg} = \frac{\as}{2\pi} \frac{e^{\gamma_E \e}}{\Gamma(1-\e)} \biggl(\frac{\mu}{\pTcut x_{qg} R_c}\biggr)^{2\e} \biggl[ \frac{1}{\e^2} C_F + \frac{1}{\e} \gamma_q + \gamma'_q + 8 A_x (R_c) C_F \biggr] \,,
\ee
and for $g\to q\bq$ we have
\be \label{eq:IC13qq}
I_C^{q\bq} = \frac{\as}{2\pi} \frac{e^{\gamma_E \e}}{\Gamma(1-\e)} \biggl(\frac{\mu}{\pTcut x_{q\bq} R_c}\biggr)^{2\e} \biggl[ \frac{1}{\e} \gamma_g^{(q)} + \gamma_g^{'\,(q)} \biggr] \,.
\ee
The above coefficients are the standard ones, 
\begin{align}
\gamma_g &= \frac{11}{6} C_A - \frac43 T_F n_f \,, & \gamma_q &= \frac32 C_F \,, \nn \\
\gamma'_g &= \biggl(\frac{67}{9} - \frac{2\pi^2}{3} \biggr) C_A - \frac{23}{9} T_F n_f \,, & \gamma'_q &= \biggl(\frac{13}{2} - \frac{2\pi^2}{3} \biggr) C_F \,.
\end{align}
The superscripts $(g)$ and $(q)$ in the integrated subtractions denote the $C_A$ and $n_f$ parts respectively.

%%%%%%%%%%%%%%%%%%%%%%%%%%%%%%%%%%%%%%%%%%%%%%%%%%%%%%%%%%%%%%%%%%%%%%%%%%%%%%%%%%%%%%%%%%%%%%%%%%%%%%%%%
\subsection{Collinear-Soft Subtractions}
\label{app:integratedcollsoft}
%%%%%%%%%%%%%%%%%%%%%%%%%%%%%%%%%%%%%%%%%%%%%%%%%%%%%%%%%%%%%%%%%%%%%%%%%%%%%%%%%%%%%%%%%%%%%%%%%%%%%%%%%

The collinear-soft subtraction $\CS_{31}$, in \eq{CSsub}, is the same as the soft subtraction $S^3_{1r}$ in \eq{S1r} with a different color factor.  Thus the integrated collinear-soft subtraction is given by the soft results above with the $R_c$ constraint and the appropriate color factor,
\be \label{eq:ICS}
I_{\CS}^{31} = \frac{\as}{2\pi} \frac{e^{\gamma_E \e}}{\Gamma(1-\e)} \biggl(\frac{\mu}{\pTcut x_c R_c}\biggr)^{2\e} (2\T_1^2) \biggl[ \frac{1}{2\e^2} + 2A_x (R_c) \biggr] \,.
\ee

%%%%%%%%%%%%%%%%%%%%%%%%%%%%%%%%%%%%%%%%%%%%%%%%%%%%%%%%%%%%%%%%%%%%%%%%%%%%%%%%%%%%%%%%%%%%%%%%%%%%%%%%%
\section{Calculation of the $\ord{\as^2}$ Clustering Logarithm}
\label{app:C2}
%%%%%%%%%%%%%%%%%%%%%%%%%%%%%%%%%%%%%%%%%%%%%%%%%%%%%%%%%%%%%%%%%%%%%%%%%%%%%%%%%%%%%%%%%%%%%%%%%%%%%%%%%

In this appendix we give the calculation of the $\ord{\as^2}$ clustering logarithm, which has been computed previously but whose elements are recycled in computing the sum of the virtual and integrated counterterm contributions to $C_3^{(2)}$.  The phase space measure is given in \eq{bornrealSn}, the matrix elements in \eq{BornME}, and the measurement function in \eq{DeltaM23}.  The matrix element and measurement function factorize into an angular part and a $p_T$-dependent part, each of which is finite and can be integrated analytically.  After a couple of simplifying steps, the result is (keeping the finite terms in $1/\eta$)
\begin{align}
\Delta S_2^{(2)} (\pTcut) &= \Bigl( \frac{\as}{\pi} \Bigr)^2 4C_A \int_0^{\infty} \df x_1 \df x_2 \, \biggl( \ln \frac{\nu}{\pTcut} - \ln (x_1 + x_2) \biggr) \frac{1}{(x_1 + x_2)^2} \biggl[ \frac{1}{2!} \langle \hat{P}_{gg} \rangle + n_f \langle \hat{P}_{q\bq} \rangle \biggr] \nn \\
& \qquad\qquad\qquad \times \bigl[ \theta(x_1 < 1) \theta(x_2 < 1) - \theta(x_1 + x_2 < 1) \bigr] \nn \\
& \qquad\qquad\qquad \times \int_{-\infty}^{\infty} \df y \int_{-\pi}^{\pi} \frac{\df\phi}{2\pi} \frac{1}{\Delta R^2} \theta(\Delta R > R) \,.
\end{align}
The $x_{1,2}$ integrals yield
\begin{align}
&\int_0^{\infty} \df x_1 \df x_2 \, \biggl( \ln \frac{\nu}{\pTcut} - \ln (x_1 + x_2) \biggr) \frac{1}{(x_1 + x_2)^2} \biggl[ \frac{1}{2!} \langle \hat{P}_{gg} \rangle + n_f \langle \hat{P}_{q\bq} \rangle \biggr] \nn \\
& \qquad \times \bigl[ \theta(x_1 < 1) \theta(x_2 < 1) - \theta(x_1 + x_2 < 1) \bigr] \nn \\
& \qquad\qquad = -\frac12 C_A \Bigl( C_2^{(1)} \ln \frac{\nu}{\pTcut} + s_2^{(1)} \Bigr) \,,
\end{align}
where $C_2^{(1)}$ is given in \eq{C21} and $s_2^{(1)}$ is a constant that was previously computed in Ref.~\cite{Stewart:2013faa}, and is equal to
\begin{align}
s_2^{(1)} &= \frac{1}{216} ( -811 + 822\ln 2 + 396 \ln^2 2 + 108 \zeta_3) + \frac{1}{108} (163 - 174\ln 2 - 72 \ln^2 2) \frac{T_F n_f}{C_A} \nn \\
&\approx 0.425 \,.
\end{align}
The angular integrals give
\be \label{eq:BornlnR}
\int_{-\infty}^{\infty} \df y \int_{-\pi}^{\pi} \frac{\df\phi}{2\pi} \frac{1}{\Delta R^2} \theta(\Delta R > R) = - \ln \frac{R}{2\pi} \,,
\ee
and hence the entire result is
\be
\Delta S_2^{(2)} (\pTcut) = \Bigl( \frac{\as C_A}{\pi} \Bigr)^2 \Bigl( C_2^{(1)} \ln \frac{\nu}{\pTcut} + s_2^{(1)} \Bigr) \ln R^2 \,.
\ee

\bibliographystyle{jhep}
\bibliography{../jets}

\end{document}